\documentclass[reprint, superscriptaddress, amsmath, amssymb, nofootinbib,onecolumn,twocolumn]{revtex4-2}

\newcommand{\cofirst}[1]{\textsuperscript{\textdaggerdbl}}
\newcommand{\equalcontribution}{%
  \vspace{1em}
  \noindent\small\textsuperscript{\textdaggerdbl} These authors contributed equally to this work.
}

\usepackage{amsmath}
\usepackage{amssymb}
\usepackage{xcolor}
\usepackage{hyperref}
\usepackage{graphicx}

\usepackage{soul}

\begin{document}

\title{Chemically Active Wetting}

\author{Susanne Liese\cofirst{1}}
\affiliation{
 Faculty of Mathematics, Natural Sciences, and Materials Engineering: Institute of Physics, University of Augsburg, Universit\"atsstra\ss e~1, 86159 Augsburg, Germany
}

\author{Xueping Zhao\cofirst{1}}
\affiliation{
Department of Mathematical Sciences,
University of Nottingham Ningbo China, Taikang East Road 199, 315100 Ningbo, China}

\author{Christoph A. Weber}
\affiliation{
 Faculty of Mathematics, Natural Sciences, and Materials Engineering: Institute of Physics, University of Augsburg, Universit\"atsstra\ss e~1, 86159 Augsburg, Germany
}
\affiliation{
corresponding authors: christoph.weber@physik.uni-augsburg.de and julicher@pks.mpg.de
}

\author{Frank Jülicher}
\affiliation{
Max Planck Institute for the Physics of Complex Systems,
 Nöthnitzer Stra\ss e~38, 01187 Dresden, Germany
}
\affiliation{
Center for Systems Biology Dresden,  Pfotenhauerstra\ss e~108, 01307 Dresden, Germany
}
\affiliation{
Cluster of Excellence Physics of Life, TU Dresden, 01062 Dresden, Germany 
}
\affiliation{
corresponding authors: christoph.weber@physik.uni-augsburg.de and julicher@pks.mpg.de
}

\begin{abstract}
Wetting of liquid droplets on passive surfaces is ubiquitous in our daily lives, and the governing physical laws are well-understood. 
When surfaces become active, however, the governing laws of wetting remain elusive. Here we propose chemically active wetting as a new class of active systems where the surface is active due to a binding process that is maintained away from equilibrium. We derive the corresponding non-equilibrium thermodynamic theory 
and show that active binding fundamentally changes the wetting behavior, leading to steady, non-equilibrium states with droplet shapes reminiscent of a pancake or a mushroom. The origin of such anomalous shapes can be explained by mapping to electrostatics, where pairs of binding sinks and sources correspond to electrostatic dipoles along the triple line. This is an example of a more general analogy, where localized chemical activity gives rise to a multipole field of the chemical potential.
The underlying physics is relevant for cells, where droplet-forming proteins can bind to membranes accompanied by the turnover of biological fuels. 
\end{abstract}

\maketitle

From water droplets spreading on glass surfaces to raindrops rolling off plant leaves, wetting phenomena are ubiquitous in our daily lives. On macroscopics scales, the laws of wetting on passive surfaces are well-understood. The shape of a wetted droplet follows a spherical cap and the contact angle between the cap and the surface is governed by the law of Young-Dupré relating the surface tensions at the triple line~\cite{young1805,dupre1869, degennes1985,gennes2004capillarity}. The stationary shape of a wetted drop can however deviate from a spherical cap in the presence of gravitation~\cite{devic2019}, visco-plasticity~\cite{martouzet2022} and heterogeneous or patterned surfaces~\cite{wu2022}.

Wetting phenomena are not limited to solid surfaces in the macroscopic world; they also manifest at mesoscopic scales on biological surfaces such as membranes. Micrometer-sized  coacervate droplets wet lipid bilayer surfaces and the contact angle follows the law of Young-Dupré~\cite{kusumaatmaja2009}. Wetting interactions on such scales can even deform membrane vesicles~\cite{liao2019,agudocanalejo2021,lu2022}, give rise to a large variety of complex droplet and vesicle shapes~\cite{mangiarotti2023} and modulate lipid packing in the membrane~\cite{mangiarotti2023}. In cells, wetting of biomolecular condensates occurs on membrane surfaces of  organelles~\cite{brangwynne2009,zhao2020, kusumaatmaja2021intracellular} and the cell's membrane~\cite{beutel2019phase, zhao2021,mangiarotti2023wetting, pombo2022wetting, sun2023assembly}. A key property of membranes is that molecules, in particular droplet components, can bind to specific receptors embedded in the membrane. In cells, binding is often active with a chemical activity that maintains binding away from equilibrium. This additional activity is typically supplied by biological fuels such as ATP or GTP~\cite{moser2009,christie2013,harrington2021}. 

Active biophysical systems exhibit a rich set of phenomena~\cite{wurtz2018, berry2018physical, weber2019, ziethen2023}. Chemically active drops can divide~\cite{zwicker2017,seyboldt2018,bauermann2022}, form liquid shells~\cite{bartolucci2021, bergmann2023liquid, bauermann2023}, and suppress coarsening~\cite{zwicker2015suppression, bartolucci2021, kirschbaum2021controlling}. 
The mismatch of chemical and phase equilibrium leads to spatial fluxes of the components even in steady state~\cite{bauermann2022chemical}. 
How fluxes that are driven by active binding processes affect wetting remains elusive. 

To understand the interplay between active binding and membrane wetting, we propose a new class of active systems, chemically active wetting, and derive the corresponding non-equilibrium thermodynamic theory.
We draw an analogy to the field of electrostatics suggesting that the triple line acts as a source multi-pole. 
The resulting fluxes deform the spherical cap-like droplet at equilibrium to shapes reminiscent of a pancake or a mushroom at non-equilibrium steady state. 

\begin{figure}[tb]
    \centering
    \includegraphics[width=7.5cm]{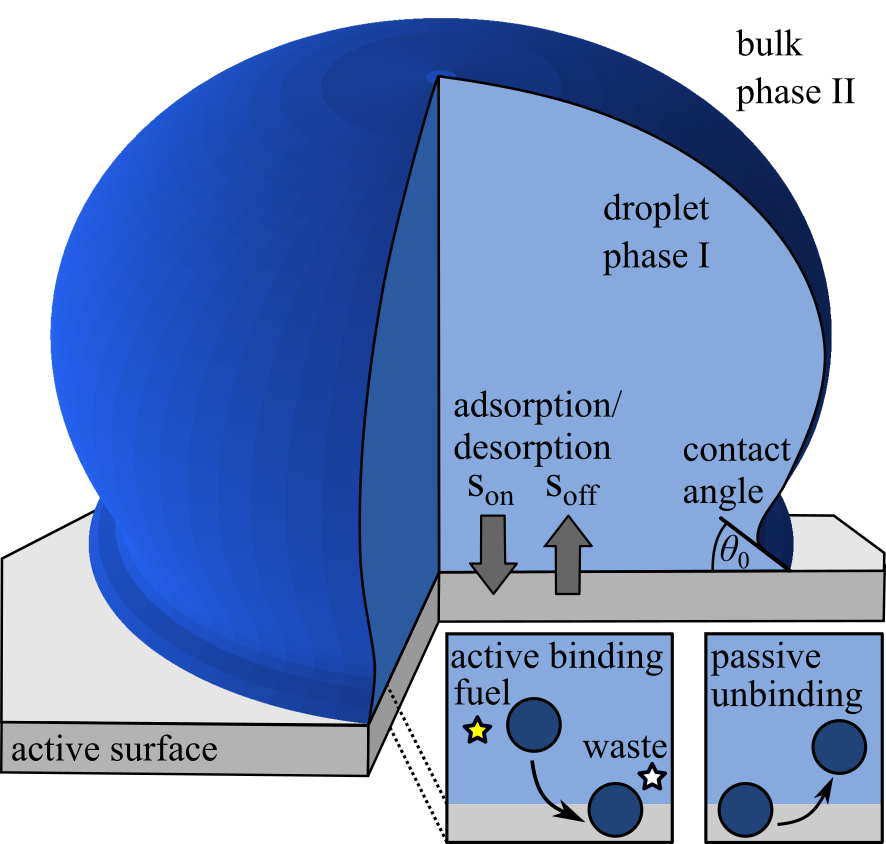}
    \caption{\textbf{Schematic illustration of wetting on chemically active surfaces.}
    The droplet can wet  the planar surface containing a droplet  (dense phase \text{I}) surrounded by a dilute phase (\text{II}). The droplet components bind to the membrane, forming a two-dimensional layer at the interface between bulk and membrane. The unbinding flux $s_{\rm off}$ is passive, while the binding flux $s_{\rm on}$ is governed by active processes, which can be realized by the consumption/production of fuel/waste.}
    \label{fig:schematic}
\end{figure}

\textbf{Theory of active wetting:} We consider a binary solute-solvent mixture that can phase-separate in the bulk and which is in contact with a membrane surface (Fig.~\ref{fig:schematic}). 
In a finite system, a droplet-phase rich in solutes  can coexist with a dilute phase in the bulk and wet the surface, enclosing a local contact angle $\theta_0$.  
Moreover, the solutes are able to bind to the membrane with a rate governed by the net desorption flux $s$ (Eq.~\eqref{eq:sbind}).
This binding process can be passive settling at binding equilibrium, or active, involving an additional external free energy $\Delta\mu_{\rm act}$.

The dynamics of active wetting can be described by a continuum theory where the fields of volume fraction in the bulk $\phi$ and the  area fraction in the membrane $\phi_{\rm m}$  are determined by the conservation laws:
\begin{subequations}\label{eq:full_dyn_model}
\label{eq:dynamics}
\begin{alignat}{2}
\label{eq:dynamics_bulk}
    \partial_t\phi  &= -\nabla\cdot\mathbf{j}  \, , \\
   \label{eq:dynamics_membrane}
    \partial_t\phi_{\rm m} &= -\nabla_{\parallel}\cdot\mathbf{j}_{\rm m} - s \, , 
\end{alignat}
where $\mathbf{j} $ and $\mathbf{j}_{\rm m}$ are to the diffusive  fluxes in bulk and  membrane. The gradient vector in the membrane plane is denoted by $\nabla_\parallel$. 

The contact angle $\theta_0$ implies a boundary condition for the bulk volume fraction $\phi$ at the membrane surface
\begin{equation}
    \label{eq:bc_wetting}
    \frac{\mathbf{n}\cdot\nabla\phi }{|\nabla\phi |}=\cos\theta_0 \, ,
\end{equation}
with the normal vector of the membrane $\mathbf{n}=(0,0,1)^{\rm T}$. The contact angle is linked to the coupling free energy $\omega$  between bulk and membrane surface via $|\nabla\phi |\sim\omega$; more details are discussed in SI, section III.B.

For solutes being conserved in membrane and bulk, the net desorption flux  $s$
is related to the normal component of diffusive bulk flux at the membrane surface:
\begin{equation}
\mathbf{n}\cdot\mathbf{j} =\frac{\nu }{\nu_{\rm m}}s \,  ,
\label{eq:boundary_condition_chem_pot}
\end{equation}
with $\nu $ and $\nu_{\rm m}$ are the molecular volume and molecular area, respectively. 

The net desorption flux is composed of the difference between an unbinding and a binding flux, $s=s_{\rm off}-s_{\rm on}$. In passive systems, the two fluxes are linked by the detailed-balance of the rates,  ${s_{\rm on}}/{s_{\rm off}}=\exp\left[-{(\mu_{\rm m}-\mu) }/{(k_{\rm B} T)}\right]$, with the chemical potentials in bulk, $\mu $, and membrane, $\mu_{\rm m}$. 
To make the surface `active',  binding is maintained away from chemical equilibrium corresponding to $\mu_{\rm m}=\mu$, we introduce an external free energy $\Delta\mu_{\rm act}$, such that 
\begin{equation}
    \frac{s_{\rm on}}{s_{\rm off}}=\exp\left[-\frac{\mu_{\rm m}-(\mu+\Delta\mu_{\rm act}) }{k_{\rm B} T}\right] \, .
\end{equation}
To ensure that this active system cannot be mapped on a passive system (i.e., by redefining the internal free energy for a constant $\Delta\mu_{\rm act}$),
the external free energy $\Delta\mu_{\rm act}$ has to be phase-dependent. For simplicity, we choose $\Delta\mu_{\rm act}=\chi_{\rm act} k_{\rm B} T\phi_0$, where $\chi_{\rm act}$ denotes the activity parameter and $\phi_0$ is the bulk volume fraction at the membrane surface. 
Note that our choice of $\Delta\mu_{\rm act}$ corresponds to a system where the magnitude of the external free energy inside the dense phase is the larger compared to the droplet surrounding.  
We consider positive and negative $\Delta\mu_{\rm act}$ which describe the tendency to enrich or deplete the membrane surface by active binding.

The diffusive fluxes in bulk and membrane,
\label{eq:fluxes}
\begin{alignat}{2}
    \mathbf{j}  &=- \Lambda (\phi ) \nabla \mu \, ,\\ \mathbf{j}_{\rm m} &= -\Lambda_{\rm m}(\phi_{\rm m}) \nabla_{\parallel} \mu_{\rm m} \, , 
\end{alignat}
\end{subequations}
are driven by gradients in the bulk and membrane chemical potential $\mu $ and $\mu_{\rm m}$, respectively, where
$\Lambda $ and $\Lambda_{\rm m}$ denote respective kinetic coefficients. Their volume and area fraction dependence is given in the methods section.

\begin{figure}[t]
    \centering
    \includegraphics[width=8cm]{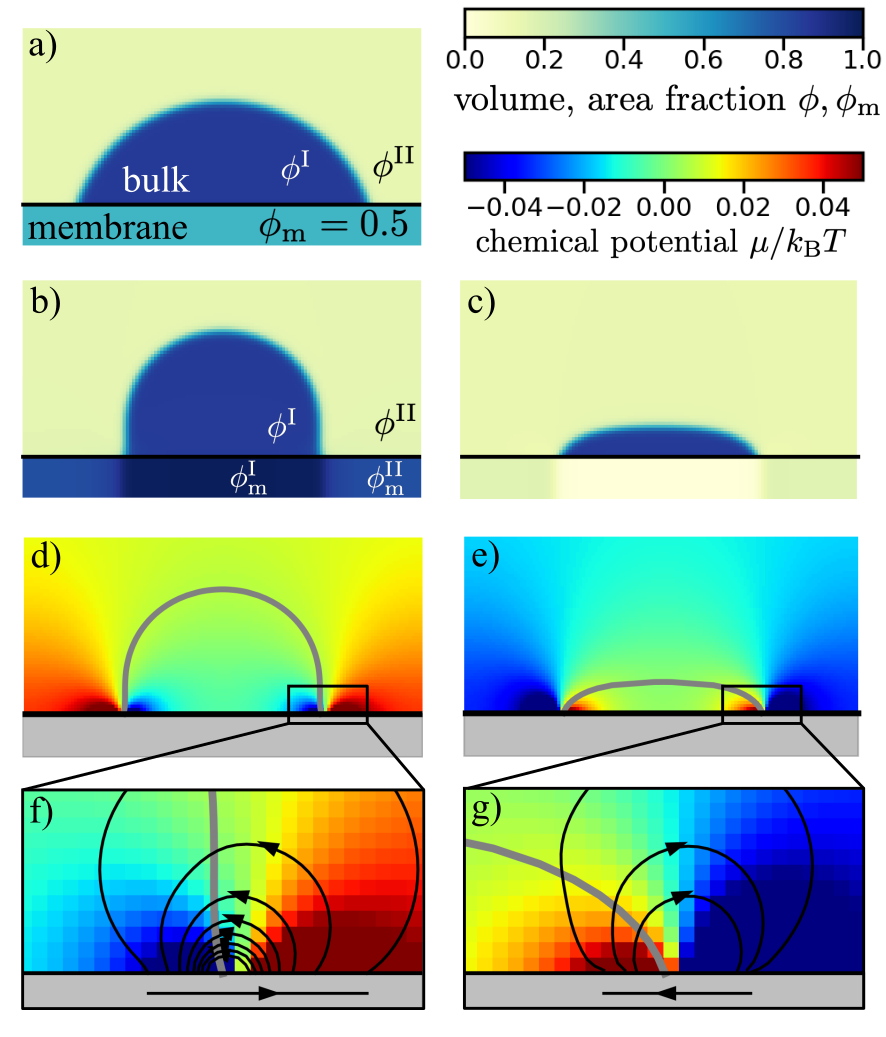}
    \caption{\textbf{Wetting on passive and chemically active surfaces:} 
    All figure panels are obtained from solving Eq.~\eqref{eq:full_dyn_model} numerically; details of the numerical method are given in SI, section II.
    \textbf{a)} Equilibrium droplet in a passive system with $\chi_{\rm act}=0$. 
    \textbf{b,c)} Stationary droplets for chemically active binding with $\chi_{\rm act}=-6$ (b) and $\chi_{\rm act}=4$ (c). 
    \textbf{d,e)} Chemical potential map that corresponds to the stationary droplets shown in subfigures b,c). The droplet shape is indicated as black line. 
    \textbf{f,g)} Chemical potential map in the vicinity of the contact line. The fluxes in bulk and in the membrane that are caused by gradients of the chemical potential are indicated by black arrows. 
    For better visibility, the membrane is shown extended in height in all subfigures. }
    \label{fig:sim_results}
\end{figure}

For a passive surface without any active binding processes ($\Delta\mu_\text{act}=0$), the steady-state solution of Eqs.~\eqref{eq:full_dyn_model} corresponds to thermodynamic equilibrium. It is characterized by a homogeneous chemical potential that is identical between bulk and membrane implying that the diffusive fluxes $\mathbf{j}$ and $\mathbf{j}_{\rm m}$, and the binding flux $s$ are each zero. In this case, the wetted droplet takes the shape of a spherical cap and the contact angle fulfills the law of Young-Dupré (Fig.~\ref{fig:sim_results}a). 

When maintaining binding away from chemical equilibrium ($\Delta\mu_\text{act}\not= 0$), we find a non-equilibrium steady state with position-dependent chemical potentials that drive diffusive fluxes in the membrane and the bulk. Such fluxes are most pronounced near the triple line. Interestingly, these localized fluxes strongly affect the shape of wetted droplets, leading to deviations from a pronounced spherical cap (Fig.~\ref{fig:sim_results}b-g). 
Depending on the value of the external free energy ($\Delta\mu_\text{act}>0$), we find shapes that are qualitatively different from a passive system with the contact line expanding or contracting relative to the passive case.  
For a chemically active surface, we observe droplet shapes that are reminiscent of a pancake or a mushroom, respectively.

\begin{figure}[b]
    \centering
    \fbox{\includegraphics[width=7.5cm]{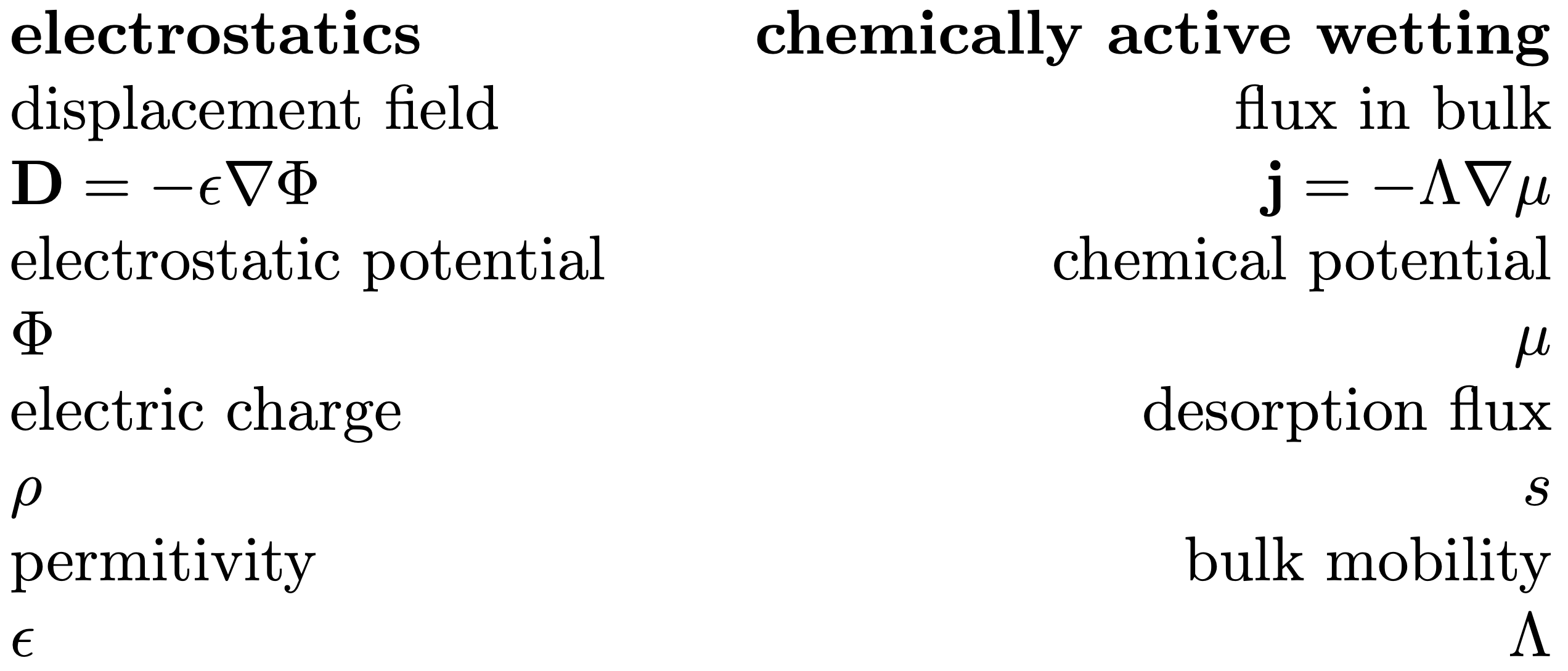}}
    \caption{\textbf{Mapping to electrostatics:} Correspondence of quantities from electrostatics (left) and wetting at active surfaces at steady state (right).}   \label{fig:analogy}
\end{figure}

\textbf{Mapping on electrostatics: }
The shape of  wetting droplets on an active surface can be understood  by drawing an analogy to electrostatics. To this end, we consider a charge free, linear dielectric medium adjacent to a non-conducting, non-polarizable medium. The interface is heterogeneously charged with a charge area density $\rho(x,y)$. According to Gauss's law, the displacement field $\mathbf{D}$ fulfills $\nabla\cdot\mathbf{D} =0$ in the absence of free charges and $\mathbf{n}\cdot\mathbf{D}=\rho$ at the interface (for more details, see SI IV). 
Comparing the electrostatic equations with the dynamic equations for active wetting Eqs.~\eqref{eq:full_dyn_model} at steady state ($\partial_t \phi =0$, $\partial_t \phi_{\rm m}=0$) suggests an mapping between electrostatics and active wetting, which is depicted in Table~\ref{fig:analogy}. 
Specifically, the net desorption flux $s$ generates a position-dependent chemical potential $\mu$ in the same way as a charge density $\rho$ gives rise to an electrostatic potential $\Phi$. Therefore, the far field of the chemical potential corresponds to the electrostatic potential field of a multipole.

To illustrate the mapping to electrostatics further, we
consider a two-dimension system for simplicity. 
In this case, a two-dimensional droplet interacts with a one-dimensional membrane or equivalently, a two-dimensional electrostatic potential resulting from a one-dimensional line charge density. The three dimensional case is discussed in the SI, section V.

The multi-pole generated by the binding flux gives rise to a chemical potential profile
which is governed to leading order by two dipoles positioned at $\pm X_p$. 
For constant mobility $\Lambda$, this chemical potential can be written as  (SI, section VA):
\begin{align}
  \mu (x,z) = &\frac{p\, \nu }{2\pi \nu_{\rm m} \Lambda }\Bigg[\frac{x-X_{\rm p}}{\left(x-X_{\rm p}\right)^2 +z ^2}\nonumber\\
  &- \frac{x+X_{\rm p}}{\left(x+X_{\rm p}\right)^2 +z ^2}\Bigg] +\mu_{\rm c} \, .
   \label{eq:chem_pot_dipole}
\end{align}
This chemical potential profile corresponds to the superposition of two dipole moments with opposite orientation and magnitude $p$. The two dipoles have a distance $2X_{\rm p}$ from each other. In Eq.~\eqref{eq:chem_pot_dipole}, $x$ and $z$ denote the lateral and horizontal coordinates and $\mu_{\rm c}$ is a constant offset, that acts as a Lagrange multiplier to ensure a symmetric droplet shape.

The chemical potential profile Eq.~\eqref{eq:chem_pot_dipole} can be derived from the multipole moments of the binding flux $s$. 
The monopole $q=\int_{-\infty}^{\infty}dx\,s(x)$ has to vanish due to particle conservation in a stationary system. 
The dipole moment of the whole active surface also vanishes due to the mirror symmetry of the droplet, which is formally written as $\int_{-\infty}^{\infty}dx\,xs(x)=0$ . Thus, the first non-vanishing moment is the quadrupole moment. 
The quadrupole moment is generated by two oppositely oriented dipoles of equal magnitude
\begin{equation}
p=\int\limits_0^\infty dx\, x s(x) \, ,
    \label{eq:dipole_moment}
\end{equation}
that are placed at $x=\pm X_{\rm p}$, with the dipole position given as
\begin{equation}
    X_{\rm p} = \frac{\int\limits_0^\infty dx\, x^2 s(x)}{2\int\limits_0^\infty dx\, x s(x)} \, .
    \label{eq:position_dipole_moment}
\end{equation}
Using the magnitude of the dipole moments and its positions, we obtain the potential profile is given in Eq.~\eqref{eq:chem_pot_dipole}.

\begin{figure}[t]
    \centering
\includegraphics[width=8cm]{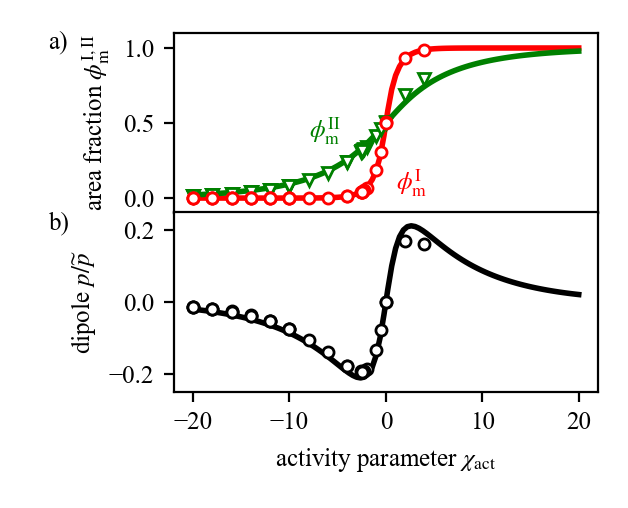} \caption{\textbf{Activity parameter affects the surface volume fractions and magnitudes of sink-source dipole:} 
    Results are obtained using the sharp interface model (solid lines) leading to Eq.~\eqref{eq:area_fraction} and Eq.~\eqref{eq:dipole_moment}, and
    continuous simulations Eqs.~\eqref{eq:full_dyn_model} (open symbols).
    \textbf{a)} The membrane area fractions $\phi_{\rm m}^\text{I}$, $\phi_{\rm m}^\text{II}$ increases with the activity parameters $\chi_\text{act}$ as it promotes binding to the surface. 
    \textbf{b)} The sink-source dipole changes sign at $\chi_\text{act}=0$ and vanishes for large $|\chi_\text{act}|$ because the active surface gets either depleted or fully occupied in both domains I and II. The dipole is scaled by $\widetilde{p}=\lambda_0^2k_0$
    }
\label{fig:sink_source_dipole}
\end{figure}

The dipole moment $p$
is caused by the mismatch of the membrane area fractions $\phi_{\rm m}^\text{I}$ and  $\phi_{\rm m}^\text{II}$ adjacent to the dense and dilute bulk phase. To estimate $\phi_{\rm m}^{\text{I},\text{II}}$, we describe the bulk droplet in the limit of a sharp interface at which local equilibrium holds leading  the dense and dilute equilibrium values $\phi^\text{I}$ and $\phi^\text{II}$. We fix the bulk chemical potential $\mu$ to be constant. 
Far from the contact line, the system becomes homogeneous even in the active case. The lateral diffusive membrane flux must therefore vanish. Subsequently the binding flux vanishes as well, which implies $s_{\rm on}=s_{\rm off}$. This results in the following relationship:
\begin{equation}
    \mu_{\rm m}-\mu-\chi_{\rm act}k_{\rm B}T\phi^{\text{I},\text{II}} = 0\, ,
    \label{eq:area_fraction}
\end{equation}
where $\mu_{\rm m}$ is a function of $\phi_{\rm m}^{\text{I},\text{II}}$. The values of $\phi_{\rm m}^{\text{I},\text{II}}$ that we determine based on Eq.~\eqref{eq:area_fraction} agree well with the simulation results (Fig.~\ref{fig:sink_source_dipole}a). 
Furthermore, using an the sharp interface  model, we find an analytic approximation for the dipole moment (see SI Sec.VI B for details)
\begin{equation}
    p=\left(D_{\rm m}^\text{I}+D_{\rm m}^\text{II}\right)\frac{\phi_{\rm m}^\text{I}-\phi_{\rm m}^\text{II}}{2} \, ,
    \label{eq:dipole_analytic}
\end{equation}
where $D_{\rm m}^{\text{I},\text{II}}$ denote the diffusion constants in a surface of area fraction $\phi_{\rm m}^{\text{I},\text{II}}$. Fig.~\ref{fig:sink_source_dipole}b) shows that the analytic results obtained from the sharp interface model agree well with the numerical solution of the continuum model (Eq.~\eqref{eq:full_dyn_model}). We see that the magnitude of $p$ exhibits a maximum around $\chi_{\rm act}=\pm2.5$ and vanishes if the magnitude of $|\chi_{\rm act}|$ increases.
The dipole vanishes for large $|\chi_{\rm act}|$ since the surface in both domains I and II gets either depleted ($\phi_m \to 0$ for negative $\chi_{\rm act}$), or fully occupied ($\phi_m \to 1$ for positive $\chi_{\rm act}$). Thus,  in both cases, the difference between $(\phi_{\rm m}^\text{I}-\phi_{\rm m}^\text{II})$ becomes small leading to a vanishing magnitude of the dipole moment according to Eq.~\eqref{eq:dipole_analytic}.

\textbf{Droplet shapes on active surfaces: }
The droplet shape is determined by the position-dependent chemical potential that  results from the active binding processes with the surface. 
Note  that there are no chemical reactions in the bulk. 
Therefore, we can consider the droplet interface between the dense droplet phase and the dilute phase to be at local equilibrium, implying a Gibbs-Thomson relation~\cite{bray1994}. For a binary mixture described by a symmetric free energy density, the mean curvature $H$ is proportional to the chemical potential, $\mu =\alpha H$, with $\alpha = \nu \gamma_0/\left(\phi ^{\text{I}}-\phi ^{\text{II}}\right)$, $\gamma_0$ the surface tension of the planar interface. Using an arc length parameterization with the arc length $S$, the mean curvature  $H=-{d\theta}/{dS}$, and $\theta$ as the angle to the horizontal $x$-axis, we find the following shape equations:
\begin{subequations}
\label{eq:shape_equations}
\begin{alignat}{3}
    \frac{d x}{dS} &= \cos\theta \, ,\\
    \frac{d z}{dS} &= \sin\theta \, ,\\
    \frac{d\theta}{dS} &=-\frac{1}{\alpha}\mu(x,z) \, ,
\end{alignat}
with the boundary conditions
\begin{alignat}{3}
    &x(0)=-X_0 \, ,\quad z(0)=0\, ,\quad \theta(0)=\theta_0 \, ,\\
    &\theta(S_{\rm mid})=0 \, ,
\end{alignat}
\end{subequations}
with $S_{\rm mid}$ denoting the mid point of the droplet interface and $X_0$ is the position of the triple line. For a given $X_0$, the offset of the chemical potential $\mu_{\rm c}$  in Eq.~\eqref{eq:chem_pot_dipole} has to be adjusted to match the boundary condition at $S_{\rm mid}$. 
We note that the area, i.e. the two-dimensional volume, can be specified instead of $X_0$. In this case, $X_0$ is a free parameter and $\mu_{\rm c}$ acts as a Lagrange multiplier of the volume.
\begin{figure}[t]
    \centering
    \includegraphics[width=8cm]{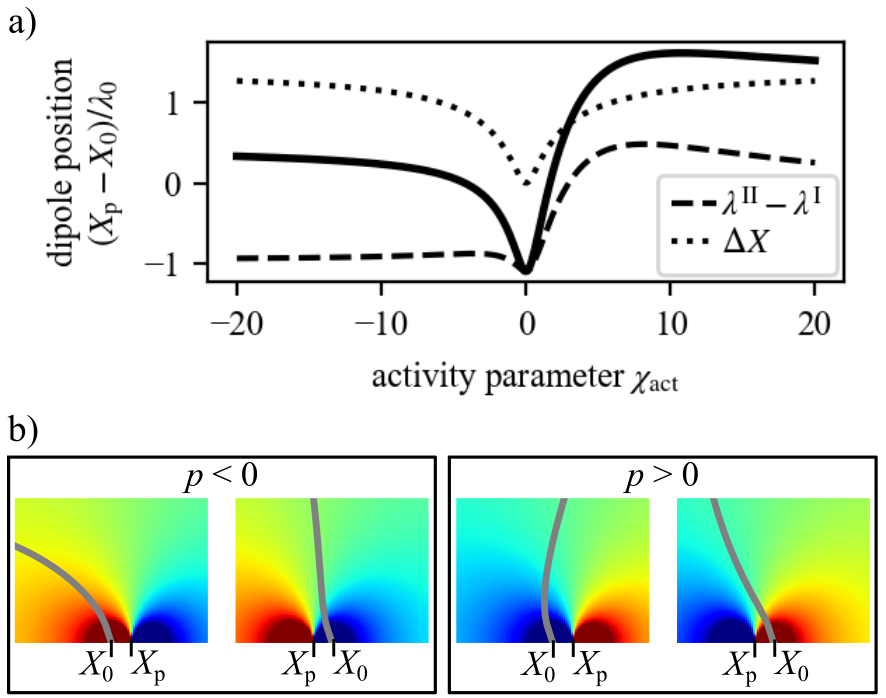}
    \caption{\textbf{Dipole position depends on activity parameter:} \textbf{a)} Using the sharp interface model, we calculate the difference between the dipole position $X_{\rm p}$ and the droplet interface at the membrane $X_0$ as a function of the activity parameter; $\lambda_0$ is the scale of the reaction-diffusion lengths. The difference  $X_{\rm p}-X_0$ consists of a symmetric term $\Delta X$ and a term that is not symmetric with respect to $\chi_{\rm act}$, due to the assymmetric contribution of the reaction-diffusion length scales.  
    \textbf{b)} Exemplary droplet shape near the triple point. We have set $\mu_{\rm c}=0$. To illustrate the impact of the dipole position, $p$ and $(X_{\rm p}-X_0)$ were set independently of each other, with $(X_{\rm p}-X_0)/\lambda_0=\pm 1$ and $p$ corresponding to $\chi_{\rm act}=\pm 0.5$.
    }
    \label{fig:dipole_position}
\end{figure}

The activity parameter $\chi_\text{act}$ affects
the position of the dipole 
\begin{equation}
    X_{\rm p} = X_0 + \Delta X + \lambda^\text{I}-\lambda^\text{II}  \, ,
\end{equation}
relative to the triple line at $X_0$ 
by a symmetric contribution $\Delta X(\chi_\text{act})=\Delta X(-\chi_\text{act})$ and an in general asymmetric contribution from the reaction-diffusion length scales $\lambda^\text{I,II}(\chi_\text{act})$
(Fig.~\ref{fig:dipole_position}a); see SI, section VIB for the expressions of $\Delta X$ and $\lambda^\text{I,II}$.
The dipole can be  deflected 
to the left or the right of the triple line.
The asymmetry of this deflection with the activity parameter results from reaction rate coefficients and diffusivities depending on volume and area fractions that vary between the domains I and II (Fig.~\ref{fig:dipole_position}a).  
The changes in the droplet position are accompanied  by pronounced changes in droplet shape in the vicinity of the triple line 
(Fig.~\ref{fig:dipole_position}b).
The shape  is calculated by solving Eqs.~\eqref{eq:shape_equations} with the chemical potential Eq.~\eqref{eq:chem_pot_dipole}; more details are given in SI, section VI.A. 
We find a rather flat, pancake-like drop with  a positive local curvature at the triple line when  $p$ and $(X_{\rm p}-X_0)$ have different signs.
Once both have the same sign, the drop has a negative curvature at the triple line leading to mushroom shapes. 

\begin{figure}[h!]
    \centering
    \includegraphics[width=8cm]{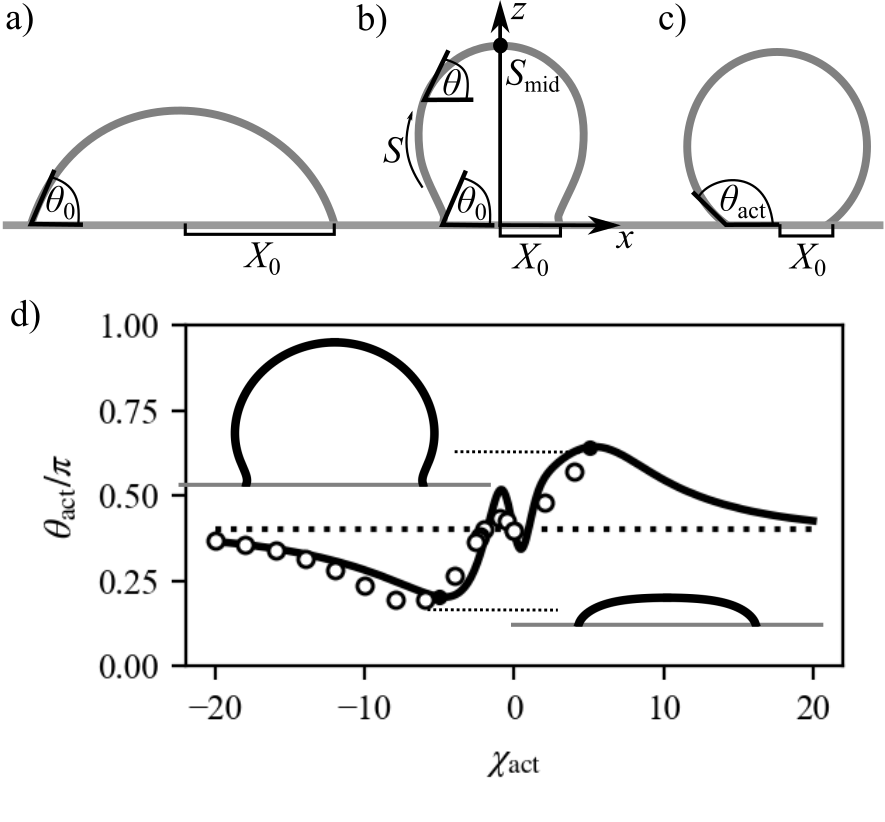}
    \caption{\textbf{Shape of droplet on active surfaces:} 
    \textbf{a)} Passive droplet. The droplet shape follows a circle segment with constant mean curvature. The contact angle is denoted as $\theta_0$.
    \textbf{b)} Active wetting. The shape of the droplet is parameterised by the arc length $S$ and the angel $\theta$ along the shape. At the membrane, the local contact angle $\theta_0$ is the same as in the passive case, even though the active contact angle, Eq.~\ref{eq:active_contact_angle}, can differ.
    \textbf{c)} A circle segment that has the same ratio of base radius squared to area as the droplet shown in subfigure b). The contact angle $\theta_{\rm act}$ of the circle segment deviates significantly from the local contact angle $\theta_0$ shown in subfigure b).
    \textbf{d)} The active contact angle depends in a non-monotonic way on the activity parameter $\chi_{\rm act}$. For  $|\chi_{\rm act}| \to \infty$, $\theta_{\rm act}$ approaches the passive value $\theta_0$, as the dipole moment vanishes. Droplet shapes are obtained using the sharp interface model (solid lines) using Eq.~\eqref{eq:shape_equations}, and
    continuous simulations Eqs.~\eqref{eq:full_dyn_model} (open symbols). Results obtained using the sharp interface model are shown for a fixed value $X_0/\lambda_0=20$. In the numerical simulations $X_0/\lambda_0$ varies between 18 and 36. The dashed line indicates the passive contact angle $\theta_0$.
    }
    \label{fig:active_contact_angle}
\end{figure}

The effects on droplet shape by the active surface can be characterized by the active contact angle $\theta_{\rm act}$:
\begin{equation}
    \frac{X_0^2}{A} = \frac{\theta_{\rm act}-\sin(\theta_{\rm act})\cos(\theta_{\rm act})}{\sin^2(\theta_{\rm act})} \, ,
    \label{eq:active_contact_angle}
\end{equation}
where $A$ is the area of the droplet, i.e., the two-dimensional equivalent of the droplet volume. The active contact angle $\theta_{\rm act}$ becomes the local contact angle $\theta_0$ when the droplet wets a passive surface leading to a circular cap shape.

The active contact angle $\theta_{\rm act}$ and thus the droplet shape is controlled  by the activity parameter $\chi_{\rm act}$.
For large and negative $\chi_{\rm act}$, $\theta_{\rm act}$ is decreased, indicating a pancake shape while for large and positive $\chi_{\rm act}$, the active contact angle in enhanced corresponding to a mushroom shape (Fig.~\ref{fig:active_contact_angle}d).
The results of the sharp interface model  (solid line) agree well with the numerical calculations for a continuous interface (open circles).  

\textbf{Active droplet wetting in experiments: }
An open question is how to experimentally realize an active system where wetting and, thereby, the droplet shape can be controlled by active binding processes to a surface. The essential ingredient is a chemical component that can form droplets and bind to a surface. Binding and unbinding need to occur in a cyclic fashion and at similar rates. Moreover, binding and unbinding have to be maintained away from chemical equilibrium, for example, via a hydrolyzing fuel component that is chemo-stated. Moreover, the binding rate coefficients $k$ need to be fast enough such that the reaction length scales $\lambda=\sqrt{D_{\rm m}/k}$ are small and localize well around the triple line. 
In this case, pronounced changes in droplet shape are expected. 
To be specific by the numbers, according to our model, pronounced shape changes occur for an activity parameter $|\chi_\text{act}|\simeq 5$
(Fig.~\eqref{fig:active_contact_angle}(d)), corresponding to a reaction-diffusion scale relative to droplet size $\lambda_0/X_0\simeq 0.05$. This case could be realized for example by a surface diffusion constant, $D_{\rm m}$, of $1 \, \mu m^2/s$  and a binding rate $k\sim k_0e^{\chi_{\rm act}}$ in the order of $1/s$.  
Thus, we propose a system that uses a ATP-driven phosphatase/kinase cycle to remove/donate a phosphate group to a phase separating component and thereby controls binding~\cite{case2019regulation}. The turn-over of ATP enables to actively regulate binding by changing the ATP concentration and thereby control the shape of wetted droplets experimentally.

\noindent\textbf{Conclusion:}
Our work shows that wetting on an active surface is significantly different from wetting on passive surfaces. We propose a novel class of active systems where an active surface that is maintained away from equilibrium by a binding process that breaks the detailed balance of the rates. We find that this binding process leads at steady-state to flux loops near the triple line. While for a passive surface, the shape of wetted droplets is a spherical cap with a minimal surface area, flux loops adjacent to active surfaces deflect the triple line where all three phases coexist. This results in droplet shapes reminiscent of a pancake or a mushroom. A striking property is that the lower dimensional active surface can strongly affect the shape of the higher dimensional droplet. 

In the quest for understanding the complexities of non-equilibrium thermodynamics, drawing analogies to the field of electrostatics revealed the underlying principles governing dynamic processes and non-equilibrium thermodynamics across diverse scales~\cite{tsori2004,deviri2021}.  In our work, we establish a conceptual mapping between non-equilibrium chemical systems and materials with electrical properties.
This conceptual mapping provides insights into the relationship between non-equilibrium chemical systems and passive materials with electrical properties.  

Our findings of shapes that significantly deviate from a spherical cap suggest that active wetting can deform and alter the structural integrity of deformable membranes.  We expect that such deformations can arise from the induced flux loops localized at the triple line acting as a local pump. 
Furthermore,  such fluxes may drive membrane shape remodeling, including changes in membrane topology. 
Such changes would provide a gateway for biomolecular transport. Wetting on active surfaces and the associated transport phenomena thus may have implications for a variety of cellular processes, including membrane budding~\cite{vutukuri2020}, and  vesicle rupture~\cite{penic2020}.

\noindent\textbf{Acknowledgments:}
We thank A.\ Honigmann and A.\ Hyman for fruitful discussions on the topics of wetting in cells and G.\ Bartolucci for helpful comments on the subject. We thank H.\ Vuijk for the critical comments on the manuscript. 
 C.\ Weber acknowledges the 
European Research Council (ERC) under the European Union’s Horizon 2020 
research and innovation programme (``Fuelled Life'' with Grant agreement No.\ 949021)
and the SPP 2191
``Molecular Mechanisms of Functional Phase Separation'' of the German Science Foundation.
F.\ J\"ulicher acknowledges funding by the Volkswagen Foundation.

\noindent\textbf{Methods:}
We describe chemically active binding such that the binding flux becomes stronger or weaker than the passive system. In contrast, the unbinding flux remains unchanged, which leads to the following representation of the net desorption flux
\begin{align}
    s=&k_0(1-\phi_{\rm m})(1-\phi_0)\times\nonumber\\
    &\left(\exp\left[\frac{\mu_{\rm m}}{k_{\rm B} T}\right]-\exp\left[\frac{\mu +\Delta\mu_{\rm act}}{k_{\rm B} T}\right]\right),
    \label{eq:sbind}
\end{align}
with $k_0$ an intrinsic binding rate, $k$ the Boltzmann constant and $T$ the temperature. In an experimental setting, our model corresponds to a scenario where fuel, which drives the active binding process, partitions into the droplet and where fuel is continuously supplied from a reservoir while waste products are cleared sufficiently fast.

To study the wetting behavior, we describe both the bulk and the membrane by a Flory-Huggins free energy density, with 
\begin{align}
    f=&\frac{k_{\rm B} T}{\nu}[ \phi\ln(\phi) + (1- \phi)\ln(1- \phi) +\chi\phi(1-\phi)]\, , 
\end{align}
in bulk and 
\begin{align}\nonumber 
    f_{\rm m}=&\frac{k_{\rm B} T}{\nu_{\rm m}}[ \phi_{\rm m}\ln(\phi_{\rm m}) + (1- \phi_{\rm m})\ln(1- \phi_{\rm m}) \\
    &\quad +\chi_{\rm m}\phi_{\rm m}(1-\phi_{\rm m})]\, ,
\end{align}
in the membrane  and  $\chi$, $\chi_{\rm m}$ the Flory-Huggins interaction parameters. 
The free energy $F$ of the system, which is composed of the bulk with volume $V$ and the membrane $m$ now reads
\begin{align}
    F&[\phi ,\phi_{\rm m}]=\int_Vd^3x\left[
    f (\phi ) +\frac{\kappa }{2}\left(\nabla\phi \right)^2\right]
    \nonumber\\
    &+\int_md^2x\left[
     f_{\rm m}(\phi_{\rm m}) +\frac{\kappa_{\rm m}}{2}\left(\nabla_\parallel\phi_{\rm m}\right)^2
     -\omega\phi_0
     \right] \, ,
     \label{eq:free_energy}
\end{align}
with $\kappa $ and $\kappa_{\rm m}$ characterising the free energy  cost for spatial inhomogeneities. The last term in Eq.~\eqref{eq:free_energy} denotes the binding energy between bulk and membrane.
For simplicity, we restrict ourselves to a coupling that is linear in the bulk volume fraction at the surface, $\phi_0=\phi (z=0)$, with a constant binding energy per unit area $\omega$. The chemical potential in bulk and membrane are obtained from the free energy as $\mu/\nu=\delta F/\delta\phi$ and $\mu_{\rm m}/\nu_{\rm m}=\delta F/\delta\phi_{\rm m}$. 
We model the mobility coefficients as $\Lambda=\Lambda^{(0)}\phi(1-\phi)$ in bulk and $\Lambda_{\rm m}=\Lambda_{\rm m}^{(0)}\phi_{\rm m}(1-\phi_{\rm m})$ in the membrane, with constant $\Lambda^{(0)}$, $\Lambda_{\rm m}^{(0)}$. Furthermore, the volume fraction $\phi$ is subjected to the boundary condition
\begin{equation}
    \mathbf{n}\cdot\nabla\phi =-\frac{\omega}{\kappa }
\end{equation}
at the the surface.\\
We define the reaction diffusion length scale $\lambda_0=\sqrt{\Lambda_{\rm m}^{(0)}k_{\rm B}T/k_0}$ as a characteristic length scale of our system.

\equalcontribution 
\bibliography{ms}

\begin{thebibliography}{42}%
\makeatletter
\providecommand \@ifxundefined [1]{%
 \@ifx{#1\undefined}
}%
\providecommand \@ifnum [1]{%
 \ifnum #1\expandafter \@firstoftwo
 \else \expandafter \@secondoftwo
 \fi
}%
\providecommand \@ifx [1]{%
 \ifx #1\expandafter \@firstoftwo
 \else \expandafter \@secondoftwo
 \fi
}%
\providecommand \natexlab [1]{#1}%
\providecommand \enquote  [1]{``#1''}%
\providecommand \bibnamefont  [1]{#1}%
\providecommand \bibfnamefont [1]{#1}%
\providecommand \citenamefont [1]{#1}%
\providecommand \href@noop [0]{\@secondoftwo}%
\providecommand \href [0]{\begingroup \@sanitize@url \@href}%
\providecommand \@href[1]{\@@startlink{#1}\@@href}%
\providecommand \@@href[1]{\endgroup#1\@@endlink}%
\providecommand \@sanitize@url [0]{\catcode `\\12\catcode `\$12\catcode
  `\&12\catcode `\#12\catcode `\^12\catcode `\_12\catcode `\%12\relax}%
\providecommand \@@startlink[1]{}%
\providecommand \@@endlink[0]{}%
\providecommand \url  [0]{\begingroup\@sanitize@url \@url }%
\providecommand \@url [1]{\endgroup\@href {#1}{\urlprefix }}%
\providecommand \urlprefix  [0]{URL }%
\providecommand \Eprint [0]{\href }%
\providecommand \doibase [0]{https://doi.org/}%
\providecommand \selectlanguage [0]{\@gobble}%
\providecommand \bibinfo  [0]{\@secondoftwo}%
\providecommand \bibfield  [0]{\@secondoftwo}%
\providecommand \translation [1]{[#1]}%
\providecommand \BibitemOpen [0]{}%
\providecommand \bibitemStop [0]{}%
\providecommand \bibitemNoStop [0]{.\EOS\space}%
\providecommand \EOS [0]{\spacefactor3000\relax}%
\providecommand \BibitemShut  [1]{\csname bibitem#1\endcsname}%
\let\auto@bib@innerbib\@empty
\bibitem [{\citenamefont {Young}(1805)}]{young1805}%
  \BibitemOpen
  \bibfield  {author} {\bibinfo {author} {\bibfnamefont {T.}~\bibnamefont
  {Young}},\ }\bibfield  {title} {\bibinfo {title} {An essay on the cohesion of
  fluids},\ }\href {https://doi.org/10.1098/rstl.1805.0005} {\bibfield
  {journal} {\bibinfo  {journal} {Phil. Trans. R. Soc.}\ }\textbf {\bibinfo
  {volume} {95}},\ \bibinfo {pages} {65} (\bibinfo {year} {1805})}\BibitemShut
  {NoStop}%
\bibitem [{\citenamefont {Dupré}\ and\ \citenamefont
  {Dupré}(1869)}]{dupre1869}%
  \BibitemOpen
  \bibfield  {author} {\bibinfo {author} {\bibfnamefont {A.}~\bibnamefont
  {Dupré}}\ and\ \bibinfo {author} {\bibfnamefont {P.}~\bibnamefont
  {Dupré}},\ }\href@noop {} {\emph {\bibinfo {title} {Théorie mécanique de
  la chaleur}}}\ (\bibinfo  {publisher} {Gauthier-Villars},\ \bibinfo {year}
  {1869})\BibitemShut {NoStop}%
\bibitem [{\citenamefont {de~Gennes}(1985)}]{degennes1985}%
  \BibitemOpen
  \bibfield  {author} {\bibinfo {author} {\bibfnamefont {P.~G.}\ \bibnamefont
  {de~Gennes}},\ }\bibfield  {title} {\bibinfo {title} {Wetting: statics and
  dynamics},\ }\href {https://doi.org/10.1103/RevModPhys.57.827} {\bibfield
  {journal} {\bibinfo  {journal} {Rev. Mod. Phys.}\ }\textbf {\bibinfo {volume}
  {57}},\ \bibinfo {pages} {827} (\bibinfo {year} {1985})}\BibitemShut
  {NoStop}%
\bibitem [{\citenamefont {Gennes}\ \emph {et~al.}(2004)\citenamefont {Gennes},
  \citenamefont {Brochard-Wyart}, \citenamefont {Qu{\'e}r{\'e}} \emph
  {et~al.}}]{gennes2004capillarity}%
  \BibitemOpen
  \bibfield  {author} {\bibinfo {author} {\bibfnamefont {P.-G.}\ \bibnamefont
  {Gennes}}, \bibinfo {author} {\bibfnamefont {F.}~\bibnamefont
  {Brochard-Wyart}}, \bibinfo {author} {\bibfnamefont {D.}~\bibnamefont
  {Qu{\'e}r{\'e}}}, \emph {et~al.},\ }\href@noop {} {\emph {\bibinfo {title}
  {Capillarity and wetting phenomena: drops, bubbles, pearls, waves}}}\
  (\bibinfo  {publisher} {Springer},\ \bibinfo {year} {2004})\BibitemShut
  {NoStop}%
\bibitem [{\citenamefont {Dević}\ \emph {et~al.}(2019)\citenamefont {Dević},
  \citenamefont {Encarnación~Escobar},\ and\ \citenamefont
  {Lohse}}]{devic2019}%
  \BibitemOpen
  \bibfield  {author} {\bibinfo {author} {\bibfnamefont {I.}~\bibnamefont
  {Dević}}, \bibinfo {author} {\bibfnamefont {J.}~\bibnamefont
  {Encarnación~Escobar}},\ and\ \bibinfo {author} {\bibfnamefont
  {D.}~\bibnamefont {Lohse}},\ }\bibfield  {title} {\bibinfo {title}
  {Equilibrium drop shapes on a tilted substrate with a chemical step},\ }\href
  {https://doi.org/10.1021/acs.langmuir.8b03557} {\bibfield  {journal}
  {\bibinfo  {journal} {Langmuir}\ }\textbf {\bibinfo {volume} {35}},\ \bibinfo
  {pages} {3880} (\bibinfo {year} {2019})}\BibitemShut {NoStop}%
\bibitem [{\citenamefont {Martouzet}\ \emph {et~al.}(2021)\citenamefont
  {Martouzet}, \citenamefont {J\o{}rgensen}, \citenamefont {Pelet},
  \citenamefont {Biance},\ and\ \citenamefont {Barentin}}]{martouzet2022}%
  \BibitemOpen
  \bibfield  {author} {\bibinfo {author} {\bibfnamefont {G.}~\bibnamefont
  {Martouzet}}, \bibinfo {author} {\bibfnamefont {L.}~\bibnamefont
  {J\o{}rgensen}}, \bibinfo {author} {\bibfnamefont {Y.}~\bibnamefont {Pelet}},
  \bibinfo {author} {\bibfnamefont {A.-L.}\ \bibnamefont {Biance}},\ and\
  \bibinfo {author} {\bibfnamefont {C.}~\bibnamefont {Barentin}},\ }\bibfield
  {title} {\bibinfo {title} {Dynamic arrest during the spreading of a yield
  stress fluid drop},\ }\href {https://doi.org/10.1103/PhysRevFluids.6.044006}
  {\bibfield  {journal} {\bibinfo  {journal} {Phys. Rev. Fluids}\ }\textbf
  {\bibinfo {volume} {6}},\ \bibinfo {pages} {044006} (\bibinfo {year}
  {2021})}\BibitemShut {NoStop}%
\bibitem [{\citenamefont {Wu}\ \emph {et~al.}(2022)\citenamefont {Wu},
  \citenamefont {Kuzina}, \citenamefont {Wang}, \citenamefont {Reischl},
  \citenamefont {Selzer}, \citenamefont {Nestler},\ and\ \citenamefont
  {Levkin}}]{wu2022}%
  \BibitemOpen
  \bibfield  {author} {\bibinfo {author} {\bibfnamefont {Y.}~\bibnamefont
  {Wu}}, \bibinfo {author} {\bibfnamefont {M.}~\bibnamefont {Kuzina}}, \bibinfo
  {author} {\bibfnamefont {F.}~\bibnamefont {Wang}}, \bibinfo {author}
  {\bibfnamefont {M.}~\bibnamefont {Reischl}}, \bibinfo {author} {\bibfnamefont
  {M.}~\bibnamefont {Selzer}}, \bibinfo {author} {\bibfnamefont
  {B.}~\bibnamefont {Nestler}},\ and\ \bibinfo {author} {\bibfnamefont {P.~A.}\
  \bibnamefont {Levkin}},\ }\bibfield  {title} {\bibinfo {title} {Equilibrium
  droplet shapes on chemically patterned surfaces: theoretical calculation,
  phase-field simulation, and experiments},\ }\href
  {https://doi.org/10.1016/j.jcis.2021.08.029} {\bibfield  {journal} {\bibinfo
  {journal} {J. Colloid Interface Sci.}\ }\textbf {\bibinfo {volume} {606}},\
  \bibinfo {pages} {1077} (\bibinfo {year} {2022})}\BibitemShut {NoStop}%
\bibitem [{\citenamefont {Kusumaatmaja}\ \emph {et~al.}(2009)\citenamefont
  {Kusumaatmaja}, \citenamefont {Li}, \citenamefont {Dimova},\ and\
  \citenamefont {Lipowsky}}]{kusumaatmaja2009}%
  \BibitemOpen
  \bibfield  {author} {\bibinfo {author} {\bibfnamefont {H.}~\bibnamefont
  {Kusumaatmaja}}, \bibinfo {author} {\bibfnamefont {Y.}~\bibnamefont {Li}},
  \bibinfo {author} {\bibfnamefont {R.}~\bibnamefont {Dimova}},\ and\ \bibinfo
  {author} {\bibfnamefont {R.}~\bibnamefont {Lipowsky}},\ }\bibfield  {title}
  {\bibinfo {title} {Intrinsic contact angle of aqueous phases at membranes and
  vesicles},\ }\href {https://doi.org/10.1103/PhysRevLett.103.238103}
  {\bibfield  {journal} {\bibinfo  {journal} {Phys. Rev. Lett.}\ }\textbf
  {\bibinfo {volume} {103}},\ \bibinfo {pages} {238103} (\bibinfo {year}
  {2009})}\BibitemShut {NoStop}%
\bibitem [{\citenamefont {Liao}\ \emph {et~al.}(2019)\citenamefont {Liao},
  \citenamefont {Fernandopulle}, \citenamefont {Wang}, \citenamefont {Choi},
  \citenamefont {Hao}, \citenamefont {Drerup}, \citenamefont {Patel},
  \citenamefont {Qamar}, \citenamefont {Nixon-Abell}, \citenamefont {Shen},
  \citenamefont {Meadows}, \citenamefont {Vendruscolo}, \citenamefont
  {Knowles}, \citenamefont {Nelson}, \citenamefont {Czekalska}, \citenamefont
  {Musteikyte}, \citenamefont {Gachechiladze}, \citenamefont {Stephens},
  \citenamefont {Pasolli}, \citenamefont {Forrest}, \citenamefont {{St
  George-Hyslop}}, \citenamefont {Lippincott-Schwartz},\ and\ \citenamefont
  {Ward}}]{liao2019}%
  \BibitemOpen
  \bibfield  {author} {\bibinfo {author} {\bibfnamefont {Y.-C.}\ \bibnamefont
  {Liao}}, \bibinfo {author} {\bibfnamefont {M.~S.}\ \bibnamefont
  {Fernandopulle}}, \bibinfo {author} {\bibfnamefont {G.}~\bibnamefont {Wang}},
  \bibinfo {author} {\bibfnamefont {H.}~\bibnamefont {Choi}}, \bibinfo {author}
  {\bibfnamefont {L.}~\bibnamefont {Hao}}, \bibinfo {author} {\bibfnamefont
  {C.~M.}\ \bibnamefont {Drerup}}, \bibinfo {author} {\bibfnamefont
  {R.}~\bibnamefont {Patel}}, \bibinfo {author} {\bibfnamefont
  {S.}~\bibnamefont {Qamar}}, \bibinfo {author} {\bibfnamefont
  {J.}~\bibnamefont {Nixon-Abell}}, \bibinfo {author} {\bibfnamefont
  {Y.}~\bibnamefont {Shen}}, \bibinfo {author} {\bibfnamefont {W.}~\bibnamefont
  {Meadows}}, \bibinfo {author} {\bibfnamefont {M.}~\bibnamefont
  {Vendruscolo}}, \bibinfo {author} {\bibfnamefont {T.~P.}\ \bibnamefont
  {Knowles}}, \bibinfo {author} {\bibfnamefont {M.}~\bibnamefont {Nelson}},
  \bibinfo {author} {\bibfnamefont {M.~A.}\ \bibnamefont {Czekalska}}, \bibinfo
  {author} {\bibfnamefont {G.}~\bibnamefont {Musteikyte}}, \bibinfo {author}
  {\bibfnamefont {M.~A.}\ \bibnamefont {Gachechiladze}}, \bibinfo {author}
  {\bibfnamefont {C.~A.}\ \bibnamefont {Stephens}}, \bibinfo {author}
  {\bibfnamefont {H.~A.}\ \bibnamefont {Pasolli}}, \bibinfo {author}
  {\bibfnamefont {L.~R.}\ \bibnamefont {Forrest}}, \bibinfo {author}
  {\bibfnamefont {P.}~\bibnamefont {{St George-Hyslop}}}, \bibinfo {author}
  {\bibfnamefont {J.}~\bibnamefont {Lippincott-Schwartz}},\ and\ \bibinfo
  {author} {\bibfnamefont {M.~E.}\ \bibnamefont {Ward}},\ }\bibfield  {title}
  {\bibinfo {title} {Rna granules hitchhike on lysosomes for long-distance
  transport, using annexin a11 as a molecular tether},\ }\href
  {https://doi.org/10.1016/j.cell.2019.08.050} {\bibfield  {journal} {\bibinfo
  {journal} {Cell}\ }\textbf {\bibinfo {volume} {179}},\ \bibinfo {pages} {147}
  (\bibinfo {year} {2019})}\BibitemShut {NoStop}%
\bibitem [{\citenamefont {Agudo-Canalejo}\ \emph {et~al.}(2021)\citenamefont
  {Agudo-Canalejo}, \citenamefont {Schultz}, \citenamefont {Chino},
  \citenamefont {Migliano}, \citenamefont {Saito}, \citenamefont
  {Koyama-Honda}, \citenamefont {Stenmark}, \citenamefont {Brech},
  \citenamefont {May}, \citenamefont {Mizushima},\ and\ \citenamefont
  {Knorr}}]{agudocanalejo2021}%
  \BibitemOpen
  \bibfield  {author} {\bibinfo {author} {\bibfnamefont {J.}~\bibnamefont
  {Agudo-Canalejo}}, \bibinfo {author} {\bibfnamefont {S.~W.}\ \bibnamefont
  {Schultz}}, \bibinfo {author} {\bibfnamefont {H.}~\bibnamefont {Chino}},
  \bibinfo {author} {\bibfnamefont {S.~M.}\ \bibnamefont {Migliano}}, \bibinfo
  {author} {\bibfnamefont {C.}~\bibnamefont {Saito}}, \bibinfo {author}
  {\bibfnamefont {I.}~\bibnamefont {Koyama-Honda}}, \bibinfo {author}
  {\bibfnamefont {H.}~\bibnamefont {Stenmark}}, \bibinfo {author}
  {\bibfnamefont {A.}~\bibnamefont {Brech}}, \bibinfo {author} {\bibfnamefont
  {A.~I.}\ \bibnamefont {May}}, \bibinfo {author} {\bibfnamefont
  {N.}~\bibnamefont {Mizushima}},\ and\ \bibinfo {author} {\bibfnamefont
  {R.~L.}\ \bibnamefont {Knorr}},\ }\bibfield  {title} {\bibinfo {title}
  {Wetting regulates autophagy of phase-separated compartments and the
  cytosol},\ }\href {https://doi.org/10.1038/s41586-020-2992-3} {\bibfield
  {journal} {\bibinfo  {journal} {Nature}\ }\textbf {\bibinfo {volume} {591}},\
  \bibinfo {pages} {142} (\bibinfo {year} {2021})}\BibitemShut {NoStop}%
\bibitem [{\citenamefont {Lu}\ \emph {et~al.}(2022)\citenamefont {Lu},
  \citenamefont {Liese}, \citenamefont {Schoenmakers}, \citenamefont {Weber},
  \citenamefont {Suzuki}, \citenamefont {Huck},\ and\ \citenamefont
  {Spruijt}}]{lu2022}%
  \BibitemOpen
  \bibfield  {author} {\bibinfo {author} {\bibfnamefont {T.}~\bibnamefont
  {Lu}}, \bibinfo {author} {\bibfnamefont {S.}~\bibnamefont {Liese}}, \bibinfo
  {author} {\bibfnamefont {L.}~\bibnamefont {Schoenmakers}}, \bibinfo {author}
  {\bibfnamefont {C.~A.}\ \bibnamefont {Weber}}, \bibinfo {author}
  {\bibfnamefont {H.}~\bibnamefont {Suzuki}}, \bibinfo {author} {\bibfnamefont
  {W.~T.~S.}\ \bibnamefont {Huck}},\ and\ \bibinfo {author} {\bibfnamefont
  {E.}~\bibnamefont {Spruijt}},\ }\bibfield  {title} {\bibinfo {title}
  {Endocytosis of coacervates into liposomes},\ }\href
  {https://doi.org/10.1021/jacs.2c04096} {\bibfield  {journal} {\bibinfo
  {journal} {J. Am. Chem. Soc.}\ }\textbf {\bibinfo {volume} {144}},\ \bibinfo
  {pages} {13451} (\bibinfo {year} {2022})}\BibitemShut {NoStop}%
\bibitem [{\citenamefont {Mangiarotti}\ \emph
  {et~al.}(2023{\natexlab{a}})\citenamefont {Mangiarotti}, \citenamefont
  {Siri}, \citenamefont {Tam}, \citenamefont {Zhao}, \citenamefont
  {Malacrida},\ and\ \citenamefont {Dimova}}]{mangiarotti2023}%
  \BibitemOpen
  \bibfield  {author} {\bibinfo {author} {\bibfnamefont {A.}~\bibnamefont
  {Mangiarotti}}, \bibinfo {author} {\bibfnamefont {M.}~\bibnamefont {Siri}},
  \bibinfo {author} {\bibfnamefont {N.}~\bibnamefont {Tam}}, \bibinfo {author}
  {\bibfnamefont {Z.}~\bibnamefont {Zhao}}, \bibinfo {author} {\bibfnamefont
  {L.}~\bibnamefont {Malacrida}},\ and\ \bibinfo {author} {\bibfnamefont
  {R.}~\bibnamefont {Dimova}},\ }\bibfield  {title} {\bibinfo {title}
  {Biomolecular condensates modulate membrane lipid packing and hydration},\
  }\href {https://doi.org/10.1038/s41467-023-41709-5} {\bibfield  {journal}
  {\bibinfo  {journal} {Nature Communications}\ }\textbf {\bibinfo {volume}
  {14}} (\bibinfo {year} {2023}{\natexlab{a}})}\BibitemShut {NoStop}%
\bibitem [{\citenamefont {Brangwynne}\ \emph {et~al.}(2009)\citenamefont
  {Brangwynne}, \citenamefont {Eckmann}, \citenamefont {Courson}, \citenamefont
  {Rybarska}, \citenamefont {Hoege}, \citenamefont {Gharakhani}, \citenamefont
  {Jülicher},\ and\ \citenamefont {Hyman}}]{brangwynne2009}%
  \BibitemOpen
  \bibfield  {author} {\bibinfo {author} {\bibfnamefont {C.~P.}\ \bibnamefont
  {Brangwynne}}, \bibinfo {author} {\bibfnamefont {C.~R.}\ \bibnamefont
  {Eckmann}}, \bibinfo {author} {\bibfnamefont {D.~S.}\ \bibnamefont
  {Courson}}, \bibinfo {author} {\bibfnamefont {A.}~\bibnamefont {Rybarska}},
  \bibinfo {author} {\bibfnamefont {C.}~\bibnamefont {Hoege}}, \bibinfo
  {author} {\bibfnamefont {J.}~\bibnamefont {Gharakhani}}, \bibinfo {author}
  {\bibfnamefont {F.}~\bibnamefont {Jülicher}},\ and\ \bibinfo {author}
  {\bibfnamefont {A.~A.}\ \bibnamefont {Hyman}},\ }\bibfield  {title} {\bibinfo
  {title} {Germline p granules are liquid droplets that localize by controlled
  dissolution/condensation},\ }\href {https://doi.org/10.1126/science.1172046}
  {\bibfield  {journal} {\bibinfo  {journal} {Science}\ }\textbf {\bibinfo
  {volume} {324}},\ \bibinfo {pages} {1729} (\bibinfo {year}
  {2009})}\BibitemShut {NoStop}%
\bibitem [{\citenamefont {Zhao}\ and\ \citenamefont {Zhang}(2020)}]{zhao2020}%
  \BibitemOpen
  \bibfield  {author} {\bibinfo {author} {\bibfnamefont {Y.}~\bibnamefont
  {Zhao}}\ and\ \bibinfo {author} {\bibfnamefont {H.}~\bibnamefont {Zhang}},\
  }\bibfield  {title} {\bibinfo {title} {Phase separation in membrane biology:
  The interplay between membrane-bound organelles and membraneless
  condensates.},\ }\href {https://doi.org/10.1016/j.devcel.2020.06.033}
  {\bibfield  {journal} {\bibinfo  {journal} {Dev Cell}\ }\textbf {\bibinfo
  {volume} {55}},\ \bibinfo {pages} {30} (\bibinfo {year} {2020})}\BibitemShut
  {NoStop}%
\bibitem [{\citenamefont {Kusumaatmaja}\ \emph {et~al.}(2021)\citenamefont
  {Kusumaatmaja}, \citenamefont {May},\ and\ \citenamefont
  {Knorr}}]{kusumaatmaja2021intracellular}%
  \BibitemOpen
  \bibfield  {author} {\bibinfo {author} {\bibfnamefont {H.}~\bibnamefont
  {Kusumaatmaja}}, \bibinfo {author} {\bibfnamefont {A.~I.}\ \bibnamefont
  {May}},\ and\ \bibinfo {author} {\bibfnamefont {R.~L.}\ \bibnamefont
  {Knorr}},\ }\bibfield  {title} {\bibinfo {title} {Intracellular wetting
  mediates contacts between liquid compartments and membrane-bound
  organelles},\ }\href@noop {} {\bibfield  {journal} {\bibinfo  {journal}
  {Journal of Cell Biology}\ }\textbf {\bibinfo {volume} {220}},\ \bibinfo
  {pages} {e202103175} (\bibinfo {year} {2021})}\BibitemShut {NoStop}%
\bibitem [{\citenamefont {Beutel}\ \emph {et~al.}(2019)\citenamefont {Beutel},
  \citenamefont {Maraspini}, \citenamefont {Pombo-Garcia}, \citenamefont
  {Martin-Lemaitre},\ and\ \citenamefont {Honigmann}}]{beutel2019phase}%
  \BibitemOpen
  \bibfield  {author} {\bibinfo {author} {\bibfnamefont {O.}~\bibnamefont
  {Beutel}}, \bibinfo {author} {\bibfnamefont {R.}~\bibnamefont {Maraspini}},
  \bibinfo {author} {\bibfnamefont {K.}~\bibnamefont {Pombo-Garcia}}, \bibinfo
  {author} {\bibfnamefont {C.}~\bibnamefont {Martin-Lemaitre}},\ and\ \bibinfo
  {author} {\bibfnamefont {A.}~\bibnamefont {Honigmann}},\ }\bibfield  {title}
  {\bibinfo {title} {Phase separation of zonula occludens proteins drives
  formation of tight junctions},\ }\href@noop {} {\bibfield  {journal}
  {\bibinfo  {journal} {Cell}\ }\textbf {\bibinfo {volume} {179}},\ \bibinfo
  {pages} {923} (\bibinfo {year} {2019})}\BibitemShut {NoStop}%
\bibitem [{\citenamefont {Zhao}\ \emph {et~al.}(2021)\citenamefont {Zhao},
  \citenamefont {Bartolucci}, \citenamefont {Honigmann}, \citenamefont
  {Jülicher},\ and\ \citenamefont {Weber}}]{zhao2021}%
  \BibitemOpen
  \bibfield  {author} {\bibinfo {author} {\bibfnamefont {X.}~\bibnamefont
  {Zhao}}, \bibinfo {author} {\bibfnamefont {G.}~\bibnamefont {Bartolucci}},
  \bibinfo {author} {\bibfnamefont {A.}~\bibnamefont {Honigmann}}, \bibinfo
  {author} {\bibfnamefont {F.}~\bibnamefont {Jülicher}},\ and\ \bibinfo
  {author} {\bibfnamefont {C.~A.}\ \bibnamefont {Weber}},\ }\bibfield  {title}
  {\bibinfo {title} {Thermodynamics of wetting, prewetting and surface phase
  transitions with surface binding},\ }\href
  {https://doi.org/10.1088/1367-2630/ac320b} {\bibfield  {journal} {\bibinfo
  {journal} {New J. Phys.}\ }\textbf {\bibinfo {volume} {23}},\ \bibinfo
  {pages} {123003} (\bibinfo {year} {2021})}\BibitemShut {NoStop}%
\bibitem [{\citenamefont {Mangiarotti}\ \emph
  {et~al.}(2023{\natexlab{b}})\citenamefont {Mangiarotti}, \citenamefont
  {Chen}, \citenamefont {Zhao}, \citenamefont {Lipowsky},\ and\ \citenamefont
  {Dimova}}]{mangiarotti2023wetting}%
  \BibitemOpen
  \bibfield  {author} {\bibinfo {author} {\bibfnamefont {A.}~\bibnamefont
  {Mangiarotti}}, \bibinfo {author} {\bibfnamefont {N.}~\bibnamefont {Chen}},
  \bibinfo {author} {\bibfnamefont {Z.}~\bibnamefont {Zhao}}, \bibinfo {author}
  {\bibfnamefont {R.}~\bibnamefont {Lipowsky}},\ and\ \bibinfo {author}
  {\bibfnamefont {R.}~\bibnamefont {Dimova}},\ }\bibfield  {title} {\bibinfo
  {title} {Wetting and complex remodeling of membranes by biomolecular
  condensates},\ }\href@noop {} {\bibfield  {journal} {\bibinfo  {journal}
  {Nature Communications}\ }\textbf {\bibinfo {volume} {14}},\ \bibinfo {pages}
  {2809} (\bibinfo {year} {2023}{\natexlab{b}})}\BibitemShut {NoStop}%
\bibitem [{\citenamefont {Pombo-Garc{\'\i}a}\ \emph {et~al.}(2022)\citenamefont
  {Pombo-Garc{\'\i}a}, \citenamefont {Martin-Lemaitre},\ and\ \citenamefont
  {Honigmann}}]{pombo2022wetting}%
  \BibitemOpen
  \bibfield  {author} {\bibinfo {author} {\bibfnamefont {K.}~\bibnamefont
  {Pombo-Garc{\'\i}a}}, \bibinfo {author} {\bibfnamefont {C.}~\bibnamefont
  {Martin-Lemaitre}},\ and\ \bibinfo {author} {\bibfnamefont {A.}~\bibnamefont
  {Honigmann}},\ }\bibfield  {title} {\bibinfo {title} {Wetting of junctional
  condensates along the apical interface promotes tight junction belt
  formation},\ }\href@noop {} {\bibfield  {journal} {\bibinfo  {journal}
  {bioRxiv}\ ,\ \bibinfo {pages} {2022}} (\bibinfo {year} {2022})}\BibitemShut
  {NoStop}%
\bibitem [{\citenamefont {Sun}\ \emph {et~al.}(2023)\citenamefont {Sun},
  \citenamefont {Zhao}, \citenamefont {Wiegand}, \citenamefont {Bartolucci},
  \citenamefont {Martin-Lemaitre}, \citenamefont {Grill}, \citenamefont
  {Hyman}, \citenamefont {Weber},\ and\ \citenamefont
  {Honigmann}}]{sun2023assembly}%
  \BibitemOpen
  \bibfield  {author} {\bibinfo {author} {\bibfnamefont {D.}~\bibnamefont
  {Sun}}, \bibinfo {author} {\bibfnamefont {X.}~\bibnamefont {Zhao}}, \bibinfo
  {author} {\bibfnamefont {T.}~\bibnamefont {Wiegand}}, \bibinfo {author}
  {\bibfnamefont {G.}~\bibnamefont {Bartolucci}}, \bibinfo {author}
  {\bibfnamefont {C.}~\bibnamefont {Martin-Lemaitre}}, \bibinfo {author}
  {\bibfnamefont {S.~W.}\ \bibnamefont {Grill}}, \bibinfo {author}
  {\bibfnamefont {A.~A.}\ \bibnamefont {Hyman}}, \bibinfo {author}
  {\bibfnamefont {C.}~\bibnamefont {Weber}},\ and\ \bibinfo {author}
  {\bibfnamefont {A.}~\bibnamefont {Honigmann}},\ }\bibfield  {title} {\bibinfo
  {title} {Assembly of tight junction belts by surface condensation and actin
  elongation},\ }\href@noop {} {\bibfield  {journal} {\bibinfo  {journal}
  {bioRxiv}\ ,\ \bibinfo {pages} {2023}} (\bibinfo {year} {2023})}\BibitemShut
  {NoStop}%
\bibitem [{\citenamefont {Moser}\ \emph {et~al.}(2009)\citenamefont {Moser},
  \citenamefont {Legate}, \citenamefont {Zent},\ and\ \citenamefont
  {Fässler}}]{moser2009}%
  \BibitemOpen
  \bibfield  {author} {\bibinfo {author} {\bibfnamefont {M.}~\bibnamefont
  {Moser}}, \bibinfo {author} {\bibfnamefont {K.~R.}\ \bibnamefont {Legate}},
  \bibinfo {author} {\bibfnamefont {R.}~\bibnamefont {Zent}},\ and\ \bibinfo
  {author} {\bibfnamefont {R.}~\bibnamefont {Fässler}},\ }\bibfield  {title}
  {\bibinfo {title} {The tail of integrins, talin, and kindlins},\ }\href
  {https://doi.org/10.1126/science.1163865} {\bibfield  {journal} {\bibinfo
  {journal} {Science}\ }\textbf {\bibinfo {volume} {324}},\ \bibinfo {pages}
  {895} (\bibinfo {year} {2009})}\BibitemShut {NoStop}%
\bibitem [{\citenamefont {Christie}\ \emph {et~al.}(2013)\citenamefont
  {Christie}, \citenamefont {Simerska}, \citenamefont {Jen}, \citenamefont
  {Jennings},\ and\ \citenamefont {Toth}}]{christie2013}%
  \BibitemOpen
  \bibfield  {author} {\bibinfo {author} {\bibfnamefont {M.~P.}\ \bibnamefont
  {Christie}}, \bibinfo {author} {\bibfnamefont {P.}~\bibnamefont {Simerska}},
  \bibinfo {author} {\bibfnamefont {F.~E.-C.}\ \bibnamefont {Jen}}, \bibinfo
  {author} {\bibfnamefont {M.~P.}\ \bibnamefont {Jennings}},\ and\ \bibinfo
  {author} {\bibfnamefont {I.}~\bibnamefont {Toth}},\ }\bibfield  {title}
  {\bibinfo {title} {Liposomes for improved enzymatic glycosylation of
  lipid-modified lactose enkephalin},\ }\href
  {https://doi.org/https://doi.org/10.1002/cplu.201300115} {\bibfield
  {journal} {\bibinfo  {journal} {ChemPlusChem}\ }\textbf {\bibinfo {volume}
  {78}},\ \bibinfo {pages} {793} (\bibinfo {year} {2013})}\BibitemShut
  {NoStop}%
\bibitem [{\citenamefont {Harrington}\ \emph {et~al.}(2021)\citenamefont
  {Harrington}, \citenamefont {Fletcher}, \citenamefont {Heermann},
  \citenamefont {Woolfson},\ and\ \citenamefont {Schwille}}]{harrington2021}%
  \BibitemOpen
  \bibfield  {author} {\bibinfo {author} {\bibfnamefont {L.}~\bibnamefont
  {Harrington}}, \bibinfo {author} {\bibfnamefont {J.~M.}\ \bibnamefont
  {Fletcher}}, \bibinfo {author} {\bibfnamefont {T.}~\bibnamefont {Heermann}},
  \bibinfo {author} {\bibfnamefont {D.~N.}\ \bibnamefont {Woolfson}},\ and\
  \bibinfo {author} {\bibfnamefont {P.}~\bibnamefont {Schwille}},\ }\bibfield
  {title} {\bibinfo {title} {De novo design of a reversible
  phosphorylation-dependent switch for membrane targeting},\ }\href
  {https://doi.org/10.1038/s41467-021-21622-5} {\bibfield  {journal} {\bibinfo
  {journal} {Nat. Commun.}\ }\textbf {\bibinfo {volume} {12}},\ \bibinfo
  {pages} {1472} (\bibinfo {year} {2021})}\BibitemShut {NoStop}%
\bibitem [{\citenamefont {Wurtz}\ and\ \citenamefont {Lee}(2018)}]{wurtz2018}%
  \BibitemOpen
  \bibfield  {author} {\bibinfo {author} {\bibfnamefont {J.~D.}\ \bibnamefont
  {Wurtz}}\ and\ \bibinfo {author} {\bibfnamefont {C.~F.}\ \bibnamefont
  {Lee}},\ }\bibfield  {title} {\bibinfo {title} {Chemical-reaction-controlled
  phase separated drops: Formation, size selection, and coarsening},\ }\href
  {https://doi.org/10.1103/PhysRevLett.120.078102} {\bibfield  {journal}
  {\bibinfo  {journal} {Phys. Rev. Lett.}\ }\textbf {\bibinfo {volume} {120}},\
  \bibinfo {pages} {078102} (\bibinfo {year} {2018})}\BibitemShut {NoStop}%
\bibitem [{\citenamefont {Berry}\ \emph {et~al.}(2018)\citenamefont {Berry},
  \citenamefont {Brangwynne},\ and\ \citenamefont
  {Haataja}}]{berry2018physical}%
  \BibitemOpen
  \bibfield  {author} {\bibinfo {author} {\bibfnamefont {J.}~\bibnamefont
  {Berry}}, \bibinfo {author} {\bibfnamefont {C.~P.}\ \bibnamefont
  {Brangwynne}},\ and\ \bibinfo {author} {\bibfnamefont {M.}~\bibnamefont
  {Haataja}},\ }\bibfield  {title} {\bibinfo {title} {Physical principles of
  intracellular organization via active and passive phase transitions},\
  }\href@noop {} {\bibfield  {journal} {\bibinfo  {journal} {Reports on
  Progress in Physics}\ }\textbf {\bibinfo {volume} {81}},\ \bibinfo {pages}
  {046601} (\bibinfo {year} {2018})}\BibitemShut {NoStop}%
\bibitem [{\citenamefont {Weber}\ \emph {et~al.}(2019)\citenamefont {Weber},
  \citenamefont {Zwicker}, \citenamefont {Jülicher},\ and\ \citenamefont
  {Lee}}]{weber2019}%
  \BibitemOpen
  \bibfield  {author} {\bibinfo {author} {\bibfnamefont {C.~A.}\ \bibnamefont
  {Weber}}, \bibinfo {author} {\bibfnamefont {D.}~\bibnamefont {Zwicker}},
  \bibinfo {author} {\bibfnamefont {F.}~\bibnamefont {Jülicher}},\ and\
  \bibinfo {author} {\bibfnamefont {C.~F.}\ \bibnamefont {Lee}},\ }\bibfield
  {title} {\bibinfo {title} {Physics of active emulsions},\ }\href
  {https://doi.org/10.1088/1361-6633/ab052b} {\bibfield  {journal} {\bibinfo
  {journal} {Rep. Prog. Phys.}\ }\textbf {\bibinfo {volume} {82}},\ \bibinfo
  {pages} {064601} (\bibinfo {year} {2019})}\BibitemShut {NoStop}%
\bibitem [{\citenamefont {Ziethen}\ \emph {et~al.}(2023)\citenamefont
  {Ziethen}, \citenamefont {Kirschbaum},\ and\ \citenamefont
  {Zwicker}}]{ziethen2023}%
  \BibitemOpen
  \bibfield  {author} {\bibinfo {author} {\bibfnamefont {N.}~\bibnamefont
  {Ziethen}}, \bibinfo {author} {\bibfnamefont {J.}~\bibnamefont
  {Kirschbaum}},\ and\ \bibinfo {author} {\bibfnamefont {D.}~\bibnamefont
  {Zwicker}},\ }\bibfield  {title} {\bibinfo {title} {Nucleation of chemically
  active droplets},\ }\href {https://doi.org/10.1103/PhysRevLett.130.248201}
  {\bibfield  {journal} {\bibinfo  {journal} {Phys. Rev. Lett.}\ }\textbf
  {\bibinfo {volume} {130}},\ \bibinfo {pages} {248201} (\bibinfo {year}
  {2023})}\BibitemShut {NoStop}%
\bibitem [{\citenamefont {Zwicker}\ \emph {et~al.}(2017)\citenamefont
  {Zwicker}, \citenamefont {Seyboldt}, \citenamefont {Weber}, \citenamefont
  {Hyman},\ and\ \citenamefont {Jülicher}}]{zwicker2017}%
  \BibitemOpen
  \bibfield  {author} {\bibinfo {author} {\bibfnamefont {D.}~\bibnamefont
  {Zwicker}}, \bibinfo {author} {\bibfnamefont {R.}~\bibnamefont {Seyboldt}},
  \bibinfo {author} {\bibfnamefont {C.~A.}\ \bibnamefont {Weber}}, \bibinfo
  {author} {\bibfnamefont {A.~A.}\ \bibnamefont {Hyman}},\ and\ \bibinfo
  {author} {\bibfnamefont {F.}~\bibnamefont {Jülicher}},\ }\bibfield  {title}
  {\bibinfo {title} {Growth and division of active droplets provides a model
  for protocells},\ }\href {https://doi.org/10.1038/nphys3984} {\bibfield
  {journal} {\bibinfo  {journal} {Nat. Phys.}\ }\textbf {\bibinfo {volume}
  {13}},\ \bibinfo {pages} {408} (\bibinfo {year} {2017})}\BibitemShut
  {NoStop}%
\bibitem [{\citenamefont {Seyboldt}\ and\ \citenamefont
  {Jülicher}(2018)}]{seyboldt2018}%
  \BibitemOpen
  \bibfield  {author} {\bibinfo {author} {\bibfnamefont {R.}~\bibnamefont
  {Seyboldt}}\ and\ \bibinfo {author} {\bibfnamefont {F.}~\bibnamefont
  {Jülicher}},\ }\bibfield  {title} {\bibinfo {title} {Role of hydrodynamic
  flows in chemically driven droplet division},\ }\href
  {https://doi.org/10.1088/1367-2630/aae735} {\bibfield  {journal} {\bibinfo
  {journal} {New J. Phys.}\ }\textbf {\bibinfo {volume} {20}},\ \bibinfo
  {pages} {105010} (\bibinfo {year} {2018})}\BibitemShut {NoStop}%
\bibitem [{\citenamefont {Bauermann}\ \emph
  {et~al.}(2022{\natexlab{a}})\citenamefont {Bauermann}, \citenamefont
  {Weber},\ and\ \citenamefont {Jülicher}}]{bauermann2022}%
  \BibitemOpen
  \bibfield  {author} {\bibinfo {author} {\bibfnamefont {J.}~\bibnamefont
  {Bauermann}}, \bibinfo {author} {\bibfnamefont {C.~A.}\ \bibnamefont
  {Weber}},\ and\ \bibinfo {author} {\bibfnamefont {F.}~\bibnamefont
  {Jülicher}},\ }\bibfield  {title} {\bibinfo {title} {Energy and matter
  supply for active droplets},\ }\href
  {https://doi.org/doi:10.1002/andp.202200132} {\bibfield  {journal} {\bibinfo
  {journal} {Ann. Phys.}\ }\textbf {\bibinfo {volume} {534}},\ \bibinfo {pages}
  {2200132} (\bibinfo {year} {2022}{\natexlab{a}})}\BibitemShut {NoStop}%
\bibitem [{\citenamefont {Bartolucci}\ \emph {et~al.}(2021)\citenamefont
  {Bartolucci}, \citenamefont {Adame-Arana}, \citenamefont {Zhao},\ and\
  \citenamefont {Weber}}]{bartolucci2021}%
  \BibitemOpen
  \bibfield  {author} {\bibinfo {author} {\bibfnamefont {G.}~\bibnamefont
  {Bartolucci}}, \bibinfo {author} {\bibfnamefont {O.}~\bibnamefont
  {Adame-Arana}}, \bibinfo {author} {\bibfnamefont {X.}~\bibnamefont {Zhao}},\
  and\ \bibinfo {author} {\bibfnamefont {C.~A.}\ \bibnamefont {Weber}},\
  }\bibfield  {title} {\bibinfo {title} {Econtrolling composition of coexisting
  phases via molecular transitions},\ }\href
  {https://doi.org/10.1016/j.bpj.2021.09.036} {\bibfield  {journal} {\bibinfo
  {journal} {Biophys. J.}\ }\textbf {\bibinfo {volume} {120}},\ \bibinfo
  {pages} {4682} (\bibinfo {year} {2021})}\BibitemShut {NoStop}%
\bibitem [{\citenamefont {Bergmann}\ \emph {et~al.}(2023)\citenamefont
  {Bergmann}, \citenamefont {Bauermann}, \citenamefont {Bartolucci},
  \citenamefont {Donau}, \citenamefont {Stasi}, \citenamefont
  {Holtmannspoetter}, \citenamefont {Julicher}, \citenamefont {Weber},\ and\
  \citenamefont {Boekhoven}}]{bergmann2023liquid}%
  \BibitemOpen
  \bibfield  {author} {\bibinfo {author} {\bibfnamefont {A.~M.}\ \bibnamefont
  {Bergmann}}, \bibinfo {author} {\bibfnamefont {J.}~\bibnamefont {Bauermann}},
  \bibinfo {author} {\bibfnamefont {G.}~\bibnamefont {Bartolucci}}, \bibinfo
  {author} {\bibfnamefont {C.}~\bibnamefont {Donau}}, \bibinfo {author}
  {\bibfnamefont {M.}~\bibnamefont {Stasi}}, \bibinfo {author} {\bibfnamefont
  {A.-L.}\ \bibnamefont {Holtmannspoetter}}, \bibinfo {author} {\bibfnamefont
  {F.}~\bibnamefont {Julicher}}, \bibinfo {author} {\bibfnamefont {C.~A.}\
  \bibnamefont {Weber}},\ and\ \bibinfo {author} {\bibfnamefont
  {J.}~\bibnamefont {Boekhoven}},\ }\bibfield  {title} {\bibinfo {title}
  {Liquid spherical shells are a non-equilibrium steady state},\ }\href@noop {}
  {\bibfield  {journal} {\bibinfo  {journal} {bioRxiv}\ ,\ \bibinfo {pages}
  {2023}} (\bibinfo {year} {2023})}\BibitemShut {NoStop}%
\bibitem [{\citenamefont {Bauermann}\ \emph {et~al.}(2023)\citenamefont
  {Bauermann}, \citenamefont {Bartolucci}, \citenamefont {Boekhoven},
  \citenamefont {Weber},\ and\ \citenamefont {Jülicher}}]{bauermann2023}%
  \BibitemOpen
  \bibfield  {author} {\bibinfo {author} {\bibfnamefont {J.}~\bibnamefont
  {Bauermann}}, \bibinfo {author} {\bibfnamefont {G.}~\bibnamefont
  {Bartolucci}}, \bibinfo {author} {\bibfnamefont {J.}~\bibnamefont
  {Boekhoven}}, \bibinfo {author} {\bibfnamefont {C.~A.}\ \bibnamefont
  {Weber}},\ and\ \bibinfo {author} {\bibfnamefont {F.}~\bibnamefont
  {Jülicher}},\ }\bibfield  {title} {\bibinfo {title} {Formation of liquid
  shells in active droplet systems}\ }\href
  {https://doi.org/10.48550/arXiv.2306.10852} {10.48550/arXiv.2306.10852}
  (\bibinfo {year} {2023})\BibitemShut {NoStop}%
\bibitem [{\citenamefont {Zwicker}\ \emph {et~al.}(2015)\citenamefont
  {Zwicker}, \citenamefont {Hyman},\ and\ \citenamefont
  {J{\"u}licher}}]{zwicker2015suppression}%
  \BibitemOpen
  \bibfield  {author} {\bibinfo {author} {\bibfnamefont {D.}~\bibnamefont
  {Zwicker}}, \bibinfo {author} {\bibfnamefont {A.~A.}\ \bibnamefont {Hyman}},\
  and\ \bibinfo {author} {\bibfnamefont {F.}~\bibnamefont {J{\"u}licher}},\
  }\bibfield  {title} {\bibinfo {title} {Suppression of ostwald ripening in
  active emulsions},\ }\href@noop {} {\bibfield  {journal} {\bibinfo  {journal}
  {Physical Review E}\ }\textbf {\bibinfo {volume} {92}},\ \bibinfo {pages}
  {012317} (\bibinfo {year} {2015})}\BibitemShut {NoStop}%
\bibitem [{\citenamefont {Kirschbaum}\ and\ \citenamefont
  {Zwicker}(2021)}]{kirschbaum2021controlling}%
  \BibitemOpen
  \bibfield  {author} {\bibinfo {author} {\bibfnamefont {J.}~\bibnamefont
  {Kirschbaum}}\ and\ \bibinfo {author} {\bibfnamefont {D.}~\bibnamefont
  {Zwicker}},\ }\bibfield  {title} {\bibinfo {title} {Controlling biomolecular
  condensates via chemical reactions},\ }\href@noop {} {\bibfield  {journal}
  {\bibinfo  {journal} {Journal of The Royal Society Interface}\ }\textbf
  {\bibinfo {volume} {18}},\ \bibinfo {pages} {20210255} (\bibinfo {year}
  {2021})}\BibitemShut {NoStop}%
\bibitem [{\citenamefont {Bauermann}\ \emph
  {et~al.}(2022{\natexlab{b}})\citenamefont {Bauermann}, \citenamefont {Laha},
  \citenamefont {McCall}, \citenamefont {Jülicher},\ and\ \citenamefont
  {Weber}}]{bauermann2022chemical}%
  \BibitemOpen
  \bibfield  {author} {\bibinfo {author} {\bibfnamefont {J.}~\bibnamefont
  {Bauermann}}, \bibinfo {author} {\bibfnamefont {S.}~\bibnamefont {Laha}},
  \bibinfo {author} {\bibfnamefont {P.~M.}\ \bibnamefont {McCall}}, \bibinfo
  {author} {\bibfnamefont {F.}~\bibnamefont {Jülicher}},\ and\ \bibinfo
  {author} {\bibfnamefont {C.~A.}\ \bibnamefont {Weber}},\ }\bibfield  {title}
  {\bibinfo {title} {Chemical kinetics and mass action in coexisting phases},\
  }\href@noop {} {\bibfield  {journal} {\bibinfo  {journal} {Journal of the
  American Chemical Society}\ }\textbf {\bibinfo {volume} {144}},\ \bibinfo
  {pages} {19294} (\bibinfo {year} {2022}{\natexlab{b}})}\BibitemShut {NoStop}%
\bibitem [{\citenamefont {Bray}(1994)}]{bray1994}%
  \BibitemOpen
  \bibfield  {author} {\bibinfo {author} {\bibfnamefont {A.}~\bibnamefont
  {Bray}},\ }\bibfield  {title} {\bibinfo {title} {Theory of phase-ordering
  kinetics},\ }\href {https://doi.org/10.1080/00018739400101505} {\bibfield
  {journal} {\bibinfo  {journal} {Advances in Physics}\ }\textbf {\bibinfo
  {volume} {43}},\ \bibinfo {pages} {357} (\bibinfo {year} {1994})},\ \Eprint
  {https://arxiv.org/abs/https://doi.org/10.1080/00018739400101505}
  {https://doi.org/10.1080/00018739400101505} \BibitemShut {NoStop}%
\bibitem [{\citenamefont {Case}\ \emph {et~al.}(2019)\citenamefont {Case},
  \citenamefont {Ditlev},\ and\ \citenamefont {Rosen}}]{case2019regulation}%
  \BibitemOpen
  \bibfield  {author} {\bibinfo {author} {\bibfnamefont {L.~B.}\ \bibnamefont
  {Case}}, \bibinfo {author} {\bibfnamefont {J.~A.}\ \bibnamefont {Ditlev}},\
  and\ \bibinfo {author} {\bibfnamefont {M.~K.}\ \bibnamefont {Rosen}},\
  }\bibfield  {title} {\bibinfo {title} {Regulation of transmembrane signaling
  by phase separation},\ }\href@noop {} {\bibfield  {journal} {\bibinfo
  {journal} {Annual review of biophysics}\ }\textbf {\bibinfo {volume} {48}},\
  \bibinfo {pages} {465} (\bibinfo {year} {2019})}\BibitemShut {NoStop}%
\bibitem [{\citenamefont {Tsori}\ and\ \citenamefont
  {de~Gennes}(2004)}]{tsori2004}%
  \BibitemOpen
  \bibfield  {author} {\bibinfo {author} {\bibfnamefont {Y.}~\bibnamefont
  {Tsori}}\ and\ \bibinfo {author} {\bibfnamefont {P.-G.}\ \bibnamefont
  {de~Gennes}},\ }\bibfield  {title} {\bibinfo {title} {Self-trapping of a
  single bacterium in its own chemoattractant},\ }\href
  {https://doi.org/10.1209/epl/i2003-10237-5} {\bibfield  {journal} {\bibinfo
  {journal} {EPL}\ }\textbf {\bibinfo {volume} {66}},\ \bibinfo {pages} {599}
  (\bibinfo {year} {2004})}\BibitemShut {NoStop}%
\bibitem [{\citenamefont {Deviri}\ and\ \citenamefont
  {Safran}(2021)}]{deviri2021}%
  \BibitemOpen
  \bibfield  {author} {\bibinfo {author} {\bibfnamefont {D.}~\bibnamefont
  {Deviri}}\ and\ \bibinfo {author} {\bibfnamefont {S.~A.}\ \bibnamefont
  {Safran}},\ }\bibfield  {title} {\bibinfo {title} {Physical theory of
  biological noise buffering by multicomponent phase separation},\ }\href
  {https://doi.org/10.1073/pnas.2100099118} {\bibfield  {journal} {\bibinfo
  {journal} {Proc. Natl. Acad. Sci. U.S.A.}\ }\textbf {\bibinfo {volume}
  {118}},\ \bibinfo {pages} {e2100099118} (\bibinfo {year} {2021})}\BibitemShut
  {NoStop}%
\bibitem [{\citenamefont {Vutukuri}\ \emph {et~al.}(2020)\citenamefont
  {Vutukuri}, \citenamefont {Hoore}, \citenamefont {Abaurrea~Velasco},
  \citenamefont {Buren}, \citenamefont {Dutto}, \citenamefont {Auth},
  \citenamefont {Fedosov}, \citenamefont {Gompper},\ and\ \citenamefont
  {Vermant}}]{vutukuri2020}%
  \BibitemOpen
  \bibfield  {author} {\bibinfo {author} {\bibfnamefont {R.}~\bibnamefont
  {Vutukuri}}, \bibinfo {author} {\bibfnamefont {M.}~\bibnamefont {Hoore}},
  \bibinfo {author} {\bibfnamefont {C.}~\bibnamefont {Abaurrea~Velasco}},
  \bibinfo {author} {\bibfnamefont {L.}~\bibnamefont {Buren}}, \bibinfo
  {author} {\bibfnamefont {A.}~\bibnamefont {Dutto}}, \bibinfo {author}
  {\bibfnamefont {T.}~\bibnamefont {Auth}}, \bibinfo {author} {\bibfnamefont
  {D.}~\bibnamefont {Fedosov}}, \bibinfo {author} {\bibfnamefont
  {G.}~\bibnamefont {Gompper}},\ and\ \bibinfo {author} {\bibfnamefont
  {J.}~\bibnamefont {Vermant}},\ }\bibfield  {title} {\bibinfo {title} {Active
  particles induce large shape deformations in giant lipid vesicles},\ }\href
  {https://doi.org/10.1038/s41586-020-2730-x} {\bibfield  {journal} {\bibinfo
  {journal} {Nature}\ }\textbf {\bibinfo {volume} {586}},\ \bibinfo {pages}
  {52} (\bibinfo {year} {2020})}\BibitemShut {NoStop}%
\bibitem [{\citenamefont {Penič}\ \emph {et~al.}(2020)\citenamefont {Penič},
  \citenamefont {Mesarec}, \citenamefont {Fošnarič}, \citenamefont
  {Mrówczyńska}, \citenamefont {Hägerstrand}, \citenamefont {Kralj-Iglič},\
  and\ \citenamefont {Iglič}}]{penic2020}%
  \BibitemOpen
  \bibfield  {author} {\bibinfo {author} {\bibfnamefont {S.}~\bibnamefont
  {Penič}}, \bibinfo {author} {\bibfnamefont {L.}~\bibnamefont {Mesarec}},
  \bibinfo {author} {\bibfnamefont {M.}~\bibnamefont {Fošnarič}}, \bibinfo
  {author} {\bibfnamefont {L.}~\bibnamefont {Mrówczyńska}}, \bibinfo {author}
  {\bibfnamefont {H.}~\bibnamefont {Hägerstrand}}, \bibinfo {author}
  {\bibfnamefont {V.}~\bibnamefont {Kralj-Iglič}},\ and\ \bibinfo {author}
  {\bibfnamefont {A.}~\bibnamefont {Iglič}},\ }\bibfield  {title} {\bibinfo
  {title} {Budding and fission of membrane vesicles: A mini review},\
  }\bibfield  {journal} {\bibinfo  {journal} {Frontiers in Physics}\ }\textbf
  {\bibinfo {volume} {8}},\ \href {https://doi.org/10.3389/fphy.2020.00342}
  {10.3389/fphy.2020.00342} (\bibinfo {year} {2020})\BibitemShut {NoStop}%
\end{thebibliography}%


\begin{thebibliography}{3}%
\makeatletter
\providecommand \@ifxundefined [1]{%
 \@ifx{#1\undefined}
}%
\providecommand \@ifnum [1]{%
 \ifnum #1\expandafter \@firstoftwo
 \else \expandafter \@secondoftwo
 \fi
}%
\providecommand \@ifx [1]{%
 \ifx #1\expandafter \@firstoftwo
 \else \expandafter \@secondoftwo
 \fi
}%
\providecommand \natexlab [1]{#1}%
\providecommand \enquote  [1]{``#1''}%
\providecommand \bibnamefont  [1]{#1}%
\providecommand \bibfnamefont [1]{#1}%
\providecommand \citenamefont [1]{#1}%
\providecommand \href@noop [0]{\@secondoftwo}%
\providecommand \href [0]{\begingroup \@sanitize@url \@href}%
\providecommand \@href[1]{\@@startlink{#1}\@@href}%
\providecommand \@@href[1]{\endgroup#1\@@endlink}%
\providecommand \@sanitize@url [0]{\catcode `\\12\catcode `\$12\catcode
  `\&12\catcode `\#12\catcode `\^12\catcode `\_12\catcode `\%12\relax}%
\providecommand \@@startlink[1]{}%
\providecommand \@@endlink[0]{}%
\providecommand \url  [0]{\begingroup\@sanitize@url \@url }%
\providecommand \@url [1]{\endgroup\@href {#1}{\urlprefix }}%
\providecommand \urlprefix  [0]{URL }%
\providecommand \Eprint [0]{\href }%
\providecommand \doibase [0]{https://doi.org/}%
\providecommand \selectlanguage [0]{\@gobble}%
\providecommand \bibinfo  [0]{\@secondoftwo}%
\providecommand \bibfield  [0]{\@secondoftwo}%
\providecommand \translation [1]{[#1]}%
\providecommand \BibitemOpen [0]{}%
\providecommand \bibitemStop [0]{}%
\providecommand \bibitemNoStop [0]{.\EOS\space}%
\providecommand \EOS [0]{\spacefactor3000\relax}%
\providecommand \BibitemShut  [1]{\csname bibitem#1\endcsname}%
\let\auto@bib@innerbib\@empty
\bibitem [{\citenamefont {Zhao}\ and\ \citenamefont
  {Wang}(2019)}]{Zhao_W2018_2}%
  \BibitemOpen
  \bibfield  {author} {\bibinfo {author} {\bibfnamefont {X.}~\bibnamefont
  {Zhao}}\ and\ \bibinfo {author} {\bibfnamefont {Q.}~\bibnamefont {Wang}},\
  }\bibfield  {title} {\bibinfo {title} {A second order fully-discrete linear
  energy stable scheme for a binary compressible viscous fluid model},\
  }\href@noop {} {\bibfield  {journal} {\bibinfo  {journal} {Journal of
  Computational Physics}\ }\textbf {\bibinfo {volume} {395}},\ \bibinfo {pages}
  {382} (\bibinfo {year} {2019})}\BibitemShut {NoStop}%
\bibitem [{\citenamefont {Zhao}\ \emph {et~al.}(2018)\citenamefont {Zhao},
  \citenamefont {Yang}, \citenamefont {Gong}, \citenamefont {Zhao},
  \citenamefont {Yang}, \citenamefont {Li},\ and\ \citenamefont
  {Wang}}]{Jia_XF_YZ_XP_XG_Jun_Q2018}%
  \BibitemOpen
  \bibfield  {author} {\bibinfo {author} {\bibfnamefont {J.}~\bibnamefont
  {Zhao}}, \bibinfo {author} {\bibfnamefont {X.}~\bibnamefont {Yang}}, \bibinfo
  {author} {\bibfnamefont {Y.}~\bibnamefont {Gong}}, \bibinfo {author}
  {\bibfnamefont {X.}~\bibnamefont {Zhao}}, \bibinfo {author} {\bibfnamefont
  {X.}~\bibnamefont {Yang}}, \bibinfo {author} {\bibfnamefont {J.}~\bibnamefont
  {Li}},\ and\ \bibinfo {author} {\bibfnamefont {Q.}~\bibnamefont {Wang}},\
  }\bibfield  {title} {\bibinfo {title} {A general strategy for numerical
  approximations of non-equilibrium models-part i: Thermodynamical systems},\
  }\href@noop {} {\bibfield  {journal} {\bibinfo  {journal} {International
  Journal of Numerical Analysis $\&$ Modeling}\ }\textbf {\bibinfo {volume}
  {15}},\ \bibinfo {pages} {884} (\bibinfo {year} {2018})}\BibitemShut
  {NoStop}%
\bibitem [{\citenamefont {Shen}\ and\ \citenamefont
  {Yang}(2010)}]{Yang_Shen_2010}%
  \BibitemOpen
  \bibfield  {author} {\bibinfo {author} {\bibfnamefont {J.}~\bibnamefont
  {Shen}}\ and\ \bibinfo {author} {\bibfnamefont {X.}~\bibnamefont {Yang}},\
  }\bibfield  {title} {\bibinfo {title} {Numerical approximations of allen-cahn
  and cahn-hilliard equations},\ }\href@noop {} {\bibfield  {journal} {\bibinfo
   {journal} {Discrete and Continuous Dynamical Systems}\ }\textbf {\bibinfo
  {volume} {28}},\ \bibinfo {pages} {1669} (\bibinfo {year}
  {2010})}\BibitemShut {NoStop}%
\end{thebibliography}%
\end{document}


\title{Supplementary Information:\\  Chemically Active Wetting
}

\author{Susanne Liese\cofirst{1}}
\affiliation{
 Faculty of Mathematics, Natural Sciences, and Materials Engineering: Institute of Physics, University of Augsburg, Universit\"atsstra\ss e~1, 86159 Augsburg, Germany
}

\author{Xueping Zhao\cofirst{1}}
\affiliation{
Department of Mathematical Sciences,
University of Nottingham Ningbo China, Taikang East Road 199, 315100 Ningbo, China}

\author{Christoph A. Weber}
\affiliation{
 Faculty of Mathematics, Natural Sciences, and Materials Engineering: Institute of Physics, University of Augsburg, Universit\"atsstra\ss e~1, 86159 Augsburg, Germany
}
\affiliation{
corresponding authors: christoph.weber@physik.uni-augsburg.de and julicher@pks.mpg.de
}

\author{Frank Jülicher}
\affiliation{
Max Planck Institute for the Physics of Complex Systems,
 Nöthnitzer Stra\ss e~38, 01187 Dresden, Germany
}
\affiliation{
Center for Systems Biology Dresden,  Pfotenhauerstra\ss e~108, 01307 Dresden, Germany
}
\affiliation{
Cluster of Excellence Physics of Life, TU Dresden, 01062 Dresden, Germany 
}
\affiliation{
corresponding authors: christoph.weber@physik.uni-augsburg.de and julicher@pks.mpg.de
}
\maketitle
\equalcontribution

\renewcommand{\theequation}{S.\arabic{equation}}
\renewcommand{\thefigure}{S.\arabic{figure}}

\section{Non-dimensionalization}
\label{sec:Non-dim}
The governing equations of the system are given as:
\begin{subequations}
\begin{align}
\partial_t \phi_{m} &=  \nabla_{||} \cdot \left( D_m \, \phi_{m} (1 - \phi_{m}) \nabla_{||} \frac{\mu_{m}}{k_\text{B}T}  \right) - s\, , \\
\partial_t \phi_{} &=  \nabla \cdot \left(D \, \phi_{} (1 - \phi_{})   \nabla \frac{\mu_{}}{k_\text{B}T} \right) \, ,
\end{align}
where $(\, \,)_{||}$ depicts the 2D differential operator. The mobilities are modeled as 
$ \Lambda_{{m}} =  \Lambda_{{m}}^0 \phi_{m} (1 - \phi_{m})$ and $ \Lambda =  \Lambda^0 \phi_{} (1 - \phi_{})$ to ensure a diffusion equation with a constant diffusion coefficient in the dilute limit, and the respective solvent, where we define the diffusion coefficient as
 $D=k_\text{B}T \Lambda^0$ and $D_{\rm m}=k_\text{B}T \Lambda_{\rm m}^0$.
The boundary conditions are given as 
\begin{align} 
0 &=   \omega +  \vect{n} \cdot \kappa_{}  \nabla \phi_{} \, ,  \quad &x \in m \, ,
\\
0 &= \vect{n} \cdot \kappa_{}  \nabla \phi_{} \, ,  \quad &x \in \partial V \, ,
\\
0 & =    \vect{t} \cdot \kappa_{m}  \nabla_\parallel \phi_{m} \,,  \quad &x \in \partial m \, ,
 \\
  -\frac{\nu}{\nu_{m} } \, s & = {\vect n} \cdot \left( D \phi_{} (1 - \phi_{})  \nabla \frac{\mu_{{}}}{k_{\text{B}}T}\right)  \quad &x \in m  \, , \\
 0 & = {\vect n} \cdot \left( D \phi_{} (1 - \phi_{})  \nabla \frac{\mu_{{}}}{k_{\text{B}}T}\right) \, ,  \quad &x \in \partial V \, .
\end{align}
\end{subequations}
We set the characteristic length scale as $l_0 = \nu^{1/3}$, and time scale as $t_0 =  \nu^{2/3}/D$. Using the rescaling 
$\tilde{\vect{x}} = \vect{x}/ l_0$ and $\tilde{t} = t \, D_b/\nu^{2/3} $, our model has non-dimensional parameters:
\begin{align}
 \tilde{D}_{\rm m} =\frac{D_{\rm m}}{D} \, , \quad
 \tilde k_0 = k_0 \nu^{2/3}/D\, , \quad \tilde{\omega}=\frac{\omega}{k_{\rm B}T}\nu^{2/3}\, ,
 \\
  \tilde{ \kappa }_{\rm m} =  \frac{1}{\nu^{2/3}} \kappa_{\rm m}  \frac{\nu_{\rm m}}{k_{\text{B}}T} \, , \quad   \tilde{ \kappa } = \kappa \frac{\nu^{1/3}}{k_{\text{B}}T}\,  .
\end{align}
Furthermore, we introduce 
\begin{equation}
    \tilde{f}=\frac{\nu}{k_{\rm B}T}f\, , \quad \tilde{f}_{\rm m}=\frac{\nu_{\rm m}}{k_{\rm B}T}f_{\rm m}\, ,
\end{equation}
the rescaled Flory-Huggings free energy density, with $f$, $f_{\rm m}$ given in Eqs. 11, 12 in the main text. And we set $\nu_{\rm m}=\nu^{2/3}$.\\
For brevity, we skip the tildes in the following. The dimensionless  equations governing of the kinetics of the system is given as:
\begin{subequations}
\label{eq:kinetic_eqns_non_dim}
\begin{align}
\partial_t \phi_{\rm m} &= \nabla_\parallel \cdot \left[  D_{\rm m}\phi_{\rm m} (1 - \phi_{\rm m}) \nabla_\parallel \left(  \frac{\partial f_{\rm m}}{\partial \phi_{\rm m}}   -  \kappa_{\rm m} \nabla_\parallel^2 \phi_{\rm m}   \right)   \right] \\
\nonumber
&\quad 
-k_0(1 - \phi_{\rm m})(1 - \phi_0)\left[
\exp\left[\frac{\partial f}{\partial \phi} - \kappa \nabla^2 \phi \right]
- \exp\left[\frac{\partial f_{\rm m}}{\partial \phi_{\rm m}} - \kappa_{\rm m} \nabla_\parallel^2 \phi_{\rm m} +\chi_{\rm act}\phi_0 \right]
\right]
\,  ,\\
\partial_t \phi_{} &= \nabla \cdot \left[  \phi_{} (1 - \phi_{}) \nabla \left(  \frac{\partial f}{\partial \phi}   -  \kappa_{} \nabla^2 \phi_{}   \right)   \right] \,  ,
\end{align}

with dimensionless boundary conditions: 
\begin{align} 
0 &=   \omega +  \vect{n} \cdot \kappa  \nabla \phi_{} \, ,  \quad &x \in m \, ,
\\
0 &=  \vect{n} \cdot \kappa  \nabla \phi_{} \, ,  \quad &x \in \partial V \, ,
\\
 0 & =      \vect{t} \cdot \kappa_{m}  \nabla_\parallel \phi_{m} \,,  \quad &x \in \partial m \, ,
 \\
 -  s & = {\vect n} \cdot \left(  \phi_{} (1 - \phi_{})  \nabla \left[\frac{\partial f}{\partial \phi}   -  \kappa_{} \nabla^2 \phi\right]\right) \,, \quad &x \in m  \, , \\
 0 & = {\vect n} \cdot \left(  \phi_{} (1 - \phi_{})  \nabla \left[\frac{\partial f}{\partial \phi}   -  \kappa_{} \nabla^2 \phi\right]\right) \, ,  \quad &x \in \partial V \, .
\end{align}
We list all the parameters and their dimensionless values in Table~\ref{tab:model_value}.

\begin{center}
\begin{table}
    \begin{tabular}{| l | c | c | }
    \hline
    \textbf{Parameter name} & \textbf{Symbol} & \textbf{rescaled value } 
    \\ \hline
     interaction coefficient in the membrane & $\chi_{m}$   & 1 \\
    \hline
     interaction coefficient in bulk & $\chi$  & 2.5 \\ 
     \hline
    binding energy per unit area& $\tilde{\omega}$ & 0.06 \\
    \hline
    diffusion coefficient in the membrane & $\tilde{D}_{\rm m}$ & 1\\
     \hline
      gradient coefficient of molecule in the membrane & $\tilde{\kappa}_m$ &   1 \\
     \hline
      gradient coefficient of molecule in the bulk & $\tilde{\kappa}$ & 1 \\
     \hline
      domain size of the bulk & $\tilde{L} \times \tilde{L}$  & 100$\times$100 \\
     \hline
    \end{tabular}
    \caption{\label{tab:model_value}\textbf{Model parameter and their dimensionless values in the model.}}
    \end{table}
\end{center}
\end{subequations}

\section{Numerical scheme of kinetic model} \label{sec:numerical_solver}

We solve the kinetic equations \ref{eq:kinetic_eqns_non_dim} with corresponding boundary conditions numerically. For the system \ref{eq:kinetic_eqns_non_dim} with the passive binding flux $s$, we initially employ the energy quadratization method \cite{Zhao_W2018_2,Jia_XF_YZ_XP_XG_Jun_Q2018} to transform the system's free energy into a quadratic formula. Subsequently, we discretize the partial differential equations using a second-order finite difference method in space and the Crank-Nicolson method in time. A stabilizing term \cite{Yang_Shen_2010} is incorporated to facilitate larger time steps. Additionally, we apply the Euler method on the exponential terms in the active binding flux.

\subsection{Numerical Evaluation of the shape equation}
To determine the drop shape based on the sharp interface model, we solve the shape equation, Eq. 7 in the main text. As the expression for the chemical potential, Eq. 2 in the main text, diverges for $x\to X_{\rm p}$, $z\to 0$, we shift the start of the integration domain by a small displacement $\Delta z$, from the surface. To optimally compare the drop shape with the numerical simulations described above, we use the mesh size $\Delta\tilde{z}=0.78$ in scaled units. This results in a shift of the starting point in the horizontal direction of $\Delta\tilde{x}=\Delta\tilde{z}/\tan(\theta_0)$ in scaled units via the local contact angle.

\section{Boundary conditions}
\subsection{Boundary condition of the flux}
\label{sec:boundary_condition_flux}
To derive the relation for the flux in bulk and the binding flux, we employ particle conservation. The total number of proteins in the system $N$ reads:
\begin{equation}
    N = \int_Vd^3x\frac{\phi}{\nu} + \int_md^2x \frac{\phi_{\rm m}}{\nu_{\rm m}},
\end{equation}
with $V$ the bulk volume and $m$ the membrane area. Particle conservation implies $dN/dt=0$, thus
\begin{equation}
    \int_Vd^3x\frac{\partial_t\phi}{\nu} = - \int_md^2x \frac{\partial_t\phi_{\rm m}}{\nu_{\rm m}}.
\end{equation}
Inserting Eqs. 1 a,b) from the main text leads to 
\begin{equation}
    \frac{1}{\nu } \int_V d^3 x \nabla\cdot\mathbf{j}  = -
        \frac{1}{\nu_{\rm m}} \int_m d^2 x \left[ \nabla_{\parallel}\cdot\mathbf{j}_{\rm m} +s\right]
\end{equation}
Applying Gauss's theorem, we obtain:
\begin{align}
    \frac{1}{\nu }\int_{\partial V}d^2x\, \mathbf{n}_{\rm V} \cdot \mathbf{j}  - \frac{1}{\nu }\int_md^2x\, \mathbf{n}\cdot \mathbf{j}  
    = -\frac{1}{\nu_{\rm m}} \int_{\partial m} dx\, \mathbf{t}\cdot \mathbf{j}_{\rm m} - \frac{1}{\nu_{\rm m}} \int_md^2x\, s,
    \label{eq:boundary_condition_chem_pot_step4}
\end{align}
with $\partial V$ the surface of the volume $V$ excluding the membrane, $\partial m$ the perimeter of the membrane surface $m$, $\mathbf{n}_{\rm V}$ and $\mathbf{t}$ the outward pointing normal vectors on $\partial V$ and $\partial m$, respectively and $\mathbf{n}=(0,0,1)^{\rm T}$ the normal vector of the membrane. In the far field gradients of the chemical potentials shall vanish. Eq. \ref{eq:boundary_condition_chem_pot_step4} simplifies to 
\begin{equation}
    \int_md^2x\, \left[ -\mathbf{n}\cdot \mathbf{j}  + \frac{\nu }{\nu_{\rm m}}s \right]=0.
    \label{eq:boundary_condition_chem_pot_step5}
\end{equation}
For the integral in  Eq. \ref{eq:boundary_condition_chem_pot_step5} to vanish for an arbitrary membrane surface $m$ the integrand has to vanish, which leads to the condition
\begin{equation}
    \mathbf{n}\cdot \mathbf{j}  = \frac{\nu }{\nu_{\rm m}}s.
\end{equation}
\subsection{Equilibrium Boundary Condition}
\label{sec:equilibrium_boundary_conditiont}
In the following the boundary condition of the volume fraction $\phi $ at the membrane interface is derived for a passive system in equilibrium. In equilibrium the functional variation of the free energy $F$, Eq. 13 in the main text is zero, with
\begin{align}
    \label{eq:si:variation}
    \delta F=&0\nonumber\\
    =&\int_Vd^3x
    \left[\left( \frac{\partial f }{\partial \phi } - \kappa \nabla^2\phi \right)\delta\phi \right]
    +\int_{\partial V}d^2x \left(\kappa \mathbf{n}_{\rm V}\cdot\nabla\phi \right)\delta\phi 
    -\int_md^2x \left(\kappa \mathbf{n}\cdot\nabla\phi \right)\delta\phi \nonumber\\
    &-\omega \int_md^2x\delta\phi 
    +\int_md^2x
    \left[\left( \frac{\partial f_{\rm m}}{\partial \phi_{\rm m}} - \kappa_{\rm m}\nabla_\parallel^2\phi_{\rm m}\right)\delta\phi_{\rm m}\right]
    +\int_{\partial m}dx\left(\mathbf{t}\cdot\nabla_\parallel\phi_{\rm m}\right)\delta\phi_{\rm m}
    .
\end{align}
For Eq. \ref{eq:si:variation} to vanish, the volume fraction has to fulfill
\begin{equation}
    \mathbf{n}\cdot\nabla\phi =-\frac{\omega}{\kappa }
\end{equation}
at the membrane interface.
 
\section{Electrostatic Analogy}
\label{sec:electrostatic_analogy}
We consider a planar two-dimensional surface with a position dependent charge density $\rho(x,y)$. Above the surface is a charge free, linear dielectric medium. The medium below the surface is non-conducting and non-polarizable. According to Gauss's law the displacement field $\mathbf{D}$ and the charge density are related as
\begin{equation}
    \nabla\cdot\mathbf{D}=\rho(x,y)\delta(z),
    \label{eq:gauss_law}
\end{equation}
where we place the charge layer at a height $z=0$. To obtain the boundary condition at the surface, we now consider a small volume with height $h$ and base area $a$. The volume is placed around the surface with the center of the volume at $z=0$. Let the area $a$ of the box be small enough so that the enclosed surface charge density can be taken as constant. Integration of Eq. \ref{eq:gauss_law} over the volume together with the Gaussian integral theorem and taking the limit $h\to 0$ leads to 
\begin{equation}
a\mathbf{n}\cdot\mathbf{D}=\rho a,
\end{equation}
which is equivalent to 
\begin{equation}
\mathbf{n}\cdot\mathbf{D}=\rho(x,y),
\end{equation}

\section{Dipole Potential in Two and Three Dimensions}
\label{sec:dipole_potential}
To derive an analytic approximation for the chemical potential in bulk, we consider an electrostatics problem in two and three dimensions. In the following, we take the dielectric constant $\epsilon$ to be space independent.

\subsection{Dipole Potential in Two Dimensions}
To determine the electrostatic potential of a dipole in two dimensions, we start with a point charge that is places at lateral position $X_{\rm p}$ and height $z=0$. The charge density $\rho$ thus reads
\begin{equation}
    \rho(x,z) = q\delta(x-X_{\rm p},z).
\end{equation}
The electrostatic potential of the charge is obtained from
\begin{equation}
    \Phi(x,z) = -\frac{1}{2\pi\epsilon}\int\int dx'dz'\rho(x',z') \ln\left(\sqrt{(x-x')^2+(z-z')^2}\right),
\end{equation}
with leads directly to 
\begin{equation}
    \Phi(x, z ) = -\frac{q}{2\pi\epsilon}\ln\left(\sqrt{(x-X_{\rm p})^2+z^2}\right)
\end{equation}
Next, we construct a dipole with dipole moment $p_{\rm q}=qd$ from two point charges placed at distances $d/2$ around the lateral positions $X_{\rm p}$. The charge density thus reads
\begin{equation}
    \rho(x,z) = 
    - q\delta\left(x-\left(X_{\rm p}-\frac{d}{2}\right),z\right)
    + q\delta\left(x-\left(X_{\rm p}+\frac{d}{2}\right),z\right).
\end{equation}
Taking the limit $d^2\ll (x-X_{\rm p})^2+z^2$ we find the dipole potential as
\begin{equation}
   \Phi(x,z) = \frac{p_{\rm q}}{2\pi\epsilon}\frac{x-X_{\rm p}}{\left(x-X_{\rm p}\right)^2 +z ^2}.
\end{equation}
Using the mapping between electrostatics and wetting at active surfaces at steady state with $\Phi\to\mu_{\rm b}$, $p_{\rm q}\to p\frac{\nu}{\nu_{\rm m}}$ and $\epsilon\to\Lambda$, the chemical potential in bulk reads
\begin{equation}
  \mu(x,z) = \frac{\bar{p}\nu}{2\pi \nu_{\rm m}\Lambda}\left[\frac{x-X_{\rm p}}{\left(x-X_{\rm p}\right)^2 +z ^2} - \frac{x+X_{\rm p}}{\left(x+X_{\rm p}\right)^2 +z ^2}\right].
\end{equation}

\subsection{Electrostatic Potential of a Dipole Ring}
We consider the three dimensional case  
with a total charge $qR_{\rm p}$ that is homogeneously distributed along a circle with radius $R_{\rm p}$ at height $z=0$. Expression the position vector $\mathbf{r}$ in cylindrical coordinates $r$, $\theta$, $z$ the charge line density $\rho(\mathbf{r}) = \frac{q}{2\pi }\delta(r-R_{\rm p},z)$. The electrostatic potential of the charge distribution is obtained from
\begin{equation}
    \Phi(\mathbf{r}) = \frac{1}{4\pi\epsilon}\int d^3r'\frac{\rho(\mathbf{r}')}{|\mathbf{r}-\mathbf{r}'|}.
\end{equation}
And the electrostatic potential
reads
\begin{equation}
    \Phi(r, \theta, z ) = \frac{qR_{\rm p}}{8\pi^2\epsilon}\int\limits_0^{2\pi}d\theta'\left[ r^2+ R_{\rm p}^2 - 2rR_{\rm p}\cos(\theta-\theta') +z^2 \right]^{-1/2}
\end{equation}
We rewrite the integrand using
\begin{equation}
    r^2+R_{\rm p}^2-2rR_{\rm p}\cos(\theta-\theta') = (r-R_{\rm p})^2 +4rR_{\rm p}\sin^2\left(\frac{\theta-\theta'}{2}\right)
\end{equation}
and substitute the integration variable by $\tau=\frac{\theta-\theta'}{2}$. Since the system is rotationally symmetric the potential does not depend on  $\theta$ and be can set without loss of generality $\theta=0$ to obtain
\begin{equation}
    \Phi(r, z ) = \frac{qR_{\rm p}}{4\pi^2\epsilon\sqrt{(R_{\rm p}-r)^2+z^2}}\int\limits_0^{\pi}d\tau \left[ 1+ \frac{4 r R_{\rm p}}{(R_{\rm p}-r)^2 +z^2}\sin^2\tau \right]^{-1/2},
\end{equation}
which leads to 
\begin{equation}
    \Phi(r, z ) = \frac{qR_{\rm p}}{2\pi^2\epsilon\sqrt{(R_{\rm p}-r)^2+z^2}}K\left(-\frac{4 r R_{\rm p}}{(R_{\rm p}-r)^2 +z^2}\right),
\end{equation}
with $K$ the complete elliptic integral of the first kind.\\
Next, we consider two charged rings at distance $d$ from each other. One ring with a charge $-qR_{\rm p}$ has radius of $R_{\rm p}-\frac{d}{2}$, while the second ring with charge $qR_{\rm p}$ has radius of $R_{\rm p}+\frac{d}{2}$. The charge density thus reads 
\begin{equation}
    \rho(\mathbf{r}) = \frac{-qR_{\rm p}}{2\pi(R_{\rm p}-d/2)}\delta(r-(R_{\rm p}-d/2),z)+\frac{qR_{\rm p}}{2\pi(R_{\rm p}+d/2)}\delta(r-(R_{\rm p}+d/2),z)
\end{equation}
and the electrostatic potential is directly obtained as 
\begin{align}
    \Phi(r, z ) = \frac{-qR_{\rm p}}{2\pi^2\epsilon\sqrt{\left(R_{\rm p}-\frac{d}{2}-r\right)^2+z^2}}K\left(-\frac{4 r \left(R_{\rm p}-\frac{d}{2}\right)}{\left(R_{\rm p}-\frac{d}{2}-r\right)^2 +z^2}\right) \\\nonumber
    + 
    \frac{qR_{\rm p}}{2\pi^2\epsilon\sqrt{\left(R_{\rm p}+\frac{d}{2}-r\right)^2+z^2}}K\left(-\frac{4 r \left(R_{\rm p}+\frac{d}{2}\right)}{\left(R_{\rm p}+\frac{d}{2}-r\right)^2 +z^2}\right).
\end{align}
To obtain the potential of a dipole ring, we take the limit $d^2\ll (r-R_{\rm p})^2+z^2$ and expand up to first order in $d/R_{\rm p}$.
\begin{equation}
     \Phi(r, z ) =\frac{q dR_{\rm p}}{2\pi}\frac{1}{\pi \epsilon}\frac{(R_{\rm p}-r)R_{\rm p}}{\left((R_{\rm p}-r)^2+z^2\right)^{3/2}}
    \left[
    \frac{\pi r R_{\rm p}}{(R_{\rm p}-r)^2+z^2}
    {}_2F_1\left(\frac{3}{2},\frac{3}{2},2,-\frac{4rR_{\rm p}}{(R_{\rm p}-r)^2+z^2}\right)
    -
    K\left(-\frac{4rR_{\rm p}}{(R_{\rm p}-r)^2+z^2}\right)
    \right],
\end{equation}
with ${}_2F_1$ the hypergeometric function. 
Using the same analogy to electrostatics as in the two-dimensional case, we directly obtain the chemical potential in bulk as 
\begin{equation}
    \mu(r,z) = \frac{p_{\rm L}\nu}{\pi\nu_{\rm m}\Lambda}\frac{(R_{\rm p}-r)R_{\rm p}}{\left((R_{\rm p}-r)^2+z^2\right)^{3/2}}
    \left[ 
     \frac{\pi rR_0}{(R_{\rm p}-r)^2+z^2} {}_2F_1\left(\frac{3}{2},\frac{3}{2},2,-\frac{4rR_{\rm p}}{(R_{\rm p}-r)^2+z^2}\right)
     -
     K\left(-\frac{4rR_{\rm p}}{(R_{\rm p}-r)^2+z^2}\right)
    \right],
\end{equation}
where the line dipole moment is defined by
\begin{equation}
    p_{\rm L}=\int\limits_0^{\infty}dr\,r^2 s(r).
\end{equation}

\subsection{Quadrupole moment}
The quadrupole moment in cartesian coordinates in two and three dimensions reads
\begin{subequations}
\begin{alignat}{2}
&\text{2D: }Q_{\rm ij}=2r_{\rm i}r_{\rm j}-r^2\delta_{\rm ij}\\
&\text{3D: }Q_{\rm ij}=3r_{\rm i}r_{\rm j}-r^2\delta_{\rm ij}.
\end{alignat}
\end{subequations}
For a charge distribution with $\rho(\mathbf{r})=\rho(r)\delta(z-0)$, \textit{i.e.} a symmetric charge distribution on a line for a two dimensional system and a rotationaly symmetric charge distribution on a plane for a three dimensional system the respective quadrupole moment reads
\begin{subequations}
\begin{alignat}{2}
&\text{2D: }
\underline{\underline{Q}} =
     \left[ {\begin{array}{cc}
    2 & 0\\
    0 & -2 \\
  \end{array} } \right]
  \int\limits_0^\infty dr r^2 \rho(r),\\
&\text{3D: }
\underline{\underline{Q}} =
     \left[ {\begin{array}{ccc}
    \pi & 0 & 0 \\
    0 & \pi & 0 \\
    0 & 0 & -2\pi \\
  \end{array} } \right]
  \int\limits_0^\infty dr r^3 \rho(r).
\end{alignat}
\end{subequations}
For the charge distributions discussed above, with
\begin{subequations}
\begin{alignat}{2}
&\text{2D: }\rho(r)=-q\delta\left(r-X_{\rm p}+\frac{d}{2}\right)+q\delta\left(r-X_{\rm p}-\frac{d}{2}\right),\\
&\text{3D: }\rho(r) = \frac{-qR_{\rm p}}{2\pi(R_{\rm p}-d/2)}\delta\left(r-R_{\rm p}+\frac{d}{2}\right)+\frac{q R_{\rm p}}{2\pi(R_{\rm p}+d/2)}\delta\left(r-R_{\rm p}-\frac{d}{2}\right)
\end{alignat}
\end{subequations}
the quadrupole moment reads
\begin{subequations}
\begin{alignat}{2}
&\text{2D: }
\underline{\underline{Q}} =
     \left[ {\begin{array}{cc}
    2 & 0\\
    0 & -2 \\
  \end{array} } \right]
  2X_{\rm p}qd\\
&\text{3D: }
\underline{\underline{Q}} =
     \left[ {\begin{array}{ccc}
    \pi & 0 & 0 \\
    0 & \pi & 0 \\
    0 & 0 & -2\pi \\
  \end{array} } \right]\frac{2qdR_{\rm p}^2}{2\pi}.
\end{alignat}
\end{subequations}
Hence, to ensure that the local dipoles exhibit the same quadrupole moment as an arbitrary symmetric (2D), or rotationaly symmetric (3D) charge distribution $\rho(r)$, $X_{\rm p}$ and $R_{\rm p}$ are set by
\begin{subequations}
\begin{alignat}{2}
&\text{2D: }X_{\rm p} = \frac{\int\limits_0^\infty dr\, r^2\rho(r)}{2\int\limits_0^\infty dr\, r\rho(r)},\\
&\text{3D: }R_{\rm p} = \frac{\int\limits_0^\infty dr\, r^3\rho(r)}{2\int\limits_0^\infty dr\, r^2\rho(r)}.
\end{alignat}
\end{subequations}

\section{Dipole moment $p$ and dipole moment position $X_{\rm p}$}
\begin{figure}[h!]
    \centering
    \includegraphics[width=7cm]{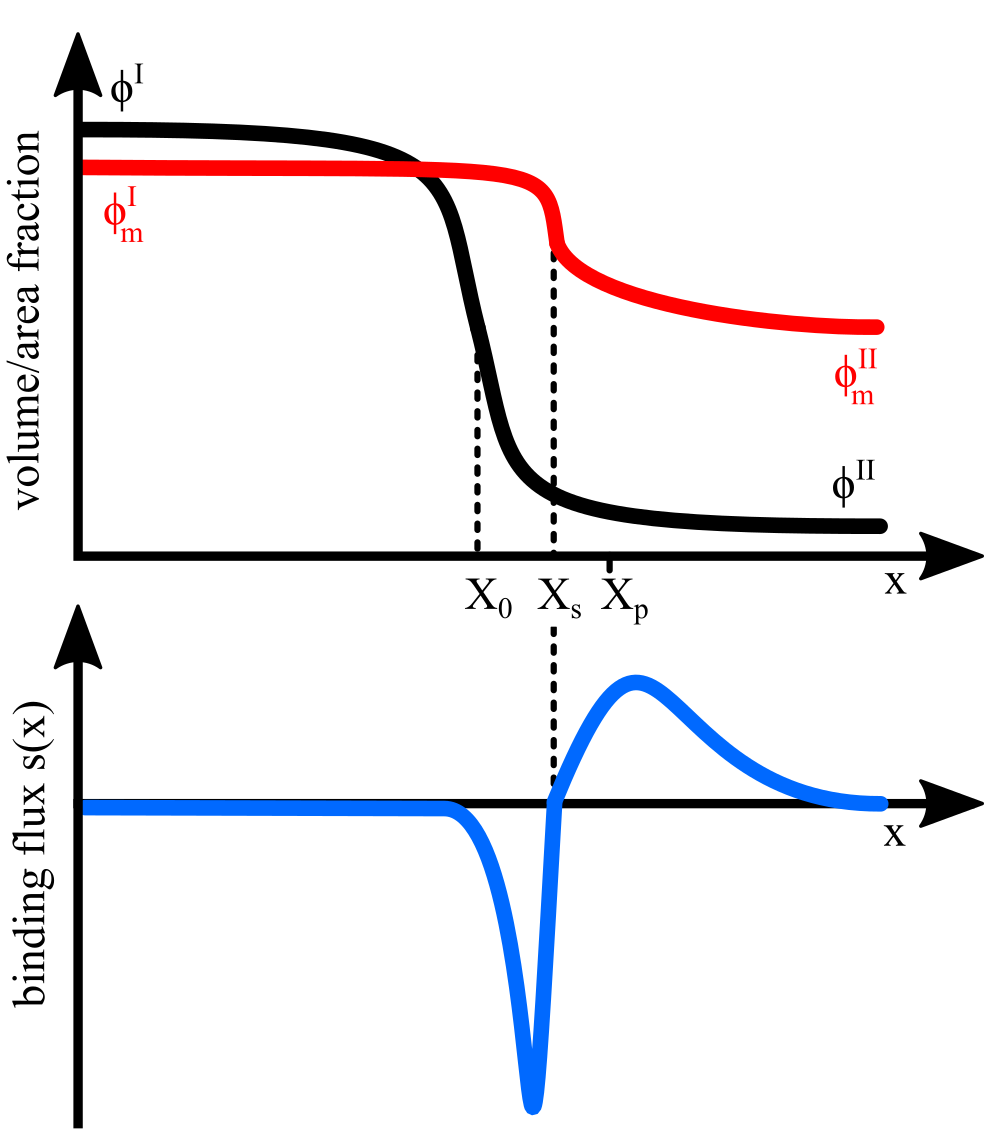}
    \caption{Schematic depiction of the vicinity of the triple point: a) In the vicinity of the triple point, the bulk volume fraction at the membrane interface (black line) transitions from the equilibrium value in the dense phase $\phi^{I}$ to the equilibrium value in the dilute phase $\phi^{II}$. The membrane area fraction (red line) transitions from $\phi_{\rm m}^{I}$ to $\phi_{\rm m}^{II}$, where the values of $\phi_{\rm m}^{I}$, $\phi_{\rm m}^{II}$ are set by the activity parameter $\chi_{\rm act}$. b) The binding flux $s$ exhibits a maximum and a minimum in the vicinity of the triple point. The magnitude of $s$ is, in general, not symmetric around $X_{\rm s}$. The schematic depiction is not drawn to scale. The characteristic features correspond to a positive $\chi_{\rm act}$.}
    \label{fig:sketch_vicinity_triple_point}
\end{figure}

In the following, we derive an expression for the dipole moment $\bar{p}$ and the dipole position $X_{\rm p}$. To this end, it is helpful first to discuss three distinct lateral positions close to the triple point $X_0$, $X_{\rm s}$ and $X_{\rm p}$, schematically depicted in Fig. \ref{fig:sketch_vicinity_triple_point}. $X_0$ denotes the position of the droplet interface on the membrane, which we define as the position where the volume fraction has the value $\phi_{\frac{1}{2}}=\left(\phi^{I}+\phi^{II}\right)/2$. $X_{\rm s}$ denotes the position where the binding flux becomes zero. Furthermore, we define $\Delta X = X_{\rm s}-X_0$.\\ 
The position of the dipole moment $X_{\rm p}$ is defined through the quadrupole moment, as discussed in the main text and further below.
Since the magnitude of the binding flux is not symmetric around $X_{\rm s}$, $X_{\rm p}$ is, in general, not equal to $X_{\rm s}$.
\subsection{Shift between $X_0$ and $X_{\rm s}$}
First, we discuss the shift between the interface position in bulk, $X_0$, and in the membrane, $X_{\rm s}$. We denote by $\phi_{\rm s}$ the bulk volume fraction adjacent to $X_{\rm s}$. For the binding flux to vanish the following condition has to be fulfilled
\begin{equation}
    \mu_{\rm m}\big|_{X_{\rm s}}-\mu-\chi_{\rm act}k_{\rm B}T\phi_{\rm s}=0.
\end{equation}
In the following, we assume that the bulk chemical potential $\mu$ can be considered constant and equal to the equilibrium value of the binodal phase separation. If gradients in the membrane area fraction are small, the membrane chemical potential is given by $\mu_{\rm m}=\nu_{\rm m}\frac{\partial f_{\rm m}}{\partial \phi_{\rm m}}$. Furthermore, we approximate the area fraction at $X_{\rm s}$ as $\phi_{\rm m}(X_{\rm s})=(\phi_{\rm m}^{I}+\phi_{\rm m}^{II})/2$, which leads to
\begin{equation}
    \phi_{\rm s}=\frac{\nu_{\rm m}\frac{\partial f_{\rm m}}{\partial \phi_{\rm m}}\big|_{\frac{\phi_{\rm m}^{I}+ \phi_{\rm m}^{II}}{2}}-\mu}{\chi_{\rm act}k_{\rm B}T}
\end{equation}
Next, we want to determine where the bulk volume fraction takes the value $\phi_{\rm s}$ relative to the position of the interface $X_0$. To this end, we recapitulate an expression for the slope $\frac{d\phi}{dx}$ at the interface between a dense and a dilute phase in equilibrium. We consider the transition from a dense to a dilute phase across a planar interface. The surface free energy $F$ reads
\begin{equation}
    F = \int_{-\infty}^\infty dx\left[f(\phi)-f(\phi^{II})-\frac{\mu}{\nu}\left(\phi-\phi^{II}\right)+\frac{\kappa}{2}\left(\frac{d\phi}{dx}\right)^2\right]
\end{equation}
with $x$ the coordinate perpendicular to the interface and where we place the interface at $x=0$. In equilibrium the variation of $F$
\begin{equation}
    \delta F = \int_{-\infty}^\infty dx\left[
    \frac{\partial f}{\partial \phi} - \frac{\mu}{\nu} -\kappa\frac{d^2\phi}{dx^2}
    \right]\delta \phi
    +\left[\kappa\frac{d\phi}{dx}\delta \phi\right]_{-\infty}^\infty
\end{equation}
has to vanish, which leads to 
\begin{equation}
    \frac{\partial f}{\partial \phi} - \frac{\mu}{\nu} -\kappa\frac{d^2\phi}{dx^2} = 0
\end{equation}
or equivalently
\begin{equation}
    \frac{d}{dx}\left(f-\frac{\mu}{\nu}\phi -\frac{\kappa}{2}\left(\frac{d\phi}{dx}\right)^2\right)=0.
\end{equation}
Since the slope $\frac{d\phi}{dx}$ has to vanish for $x\to\infty$, we find
\begin{equation}
    \frac{d\phi}{dx}=-\sqrt{
    \frac{2}{\kappa}
    \left( f(\phi) - f(\phi^{II}) -\frac{\mu}{\nu}(\phi-\phi^{II})
    \right)}.
\end{equation}
The slope at the interface $X_0$ thus reads
\begin{equation}
    \frac{d\phi}{dx}\bigg|_{X_0}=-\sqrt{
    \frac{2}{\kappa}
    \left( f(\phi_{\frac{1}{2}}) - f(\phi^{II}) -\frac{\mu}{\nu}(\phi_{\frac{1}{2}}-\phi^{II})
    \right)}.
\end{equation}
As a last step, we assume that the slope $\frac{d\phi}{dx}$ in the interface region can be considered constant, which leads to the relation
\begin{equation}
    \frac{\phi_{\rm s}-\phi_{\frac{1}{2}}}{X_{\rm s}-X_0}=\frac{d\phi}{dx}\bigg|_{X_0}.
\end{equation}
The shift between $X_0$ and $X_{\rm s}$, with $\Delta X = X_{\rm s}-X_0$ becomes
\begin{equation}
    \Delta X=\frac{ \frac{\phi^{I}+\phi^{II}}{2} - \nicefrac{\left(\frac{\nu_{\rm m}}{k_{\rm B}T}\frac{\partial f_{\rm m}}{\partial \phi_{\rm m}}\big|_{\frac{\phi_{\rm m}^{I}+\phi_{\rm m}^{II}}{2}}-\frac{\mu}{k_{\rm B}T}\right)}{\chi_{\rm act}}}{\sqrt{
    \frac{2}{\kappa}
    \left( f(\phi_{\frac{1}{2}}) - f(\phi^{II}) -\frac{\mu}{\nu}(\phi_{\frac{1}{2}}-\phi^{II})
    \right)}}.
    \label{eq:si:shift_Xs-X0}
\end{equation}
\subsection{Two-dimensional system}
\begin{figure}[h!]
    \centering
    \includegraphics[width=7cm]{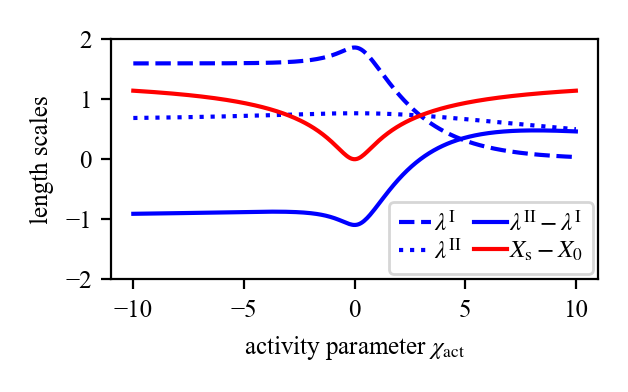}
    \caption{Relevant length scales in connection to the dipole position: While the shift between the membrane interface and the droplet interface $X_{\rm s}-X_0$ is symmetric with $\chi_{\rm act}$, neither $\lambda^{I}$ nor $\lambda^{II}$ exhibit such a symmetry. All lengths are given in units of the reaction diffusion length $\lambda_0$.}
    \label{fig:length_scales}
\end{figure}

For the stationary solution, the binding flux $s$ and the lateral membrane flux $j_{\rm m}$ are related as 
\begin{equation}
    s=-\frac{d}{dx}j_{\rm m}.
    \label{eq:si:stationary}
\end{equation} 
Linearizing $j_{\rm m}$ around $\phi_{\rm m}^{I}$ and $\phi_{\rm m}^{II}$ respectively, leads to
\begin{equation}
    j_{\rm m}=\begin{cases}
    -D_{\rm m}^{I}\frac{d}{dx}\phi_{\rm m},\quad x<X_{\rm s}\\
    -D_{\rm m}^{II}\frac{d}{dx}\phi_{\rm m},\quad x>X_{\rm s},
    \end{cases}
    \label{eq:si:dipole_membrane_flux_cases}
\end{equation}
with $D_{\rm m}^{I,II}=D_{\rm m}^{(0)}\phi_{\rm m}^{I,II}\left(1-\phi_{\rm m}^{I,II}\right)\frac{\nu_{\rm m}}{k_{\rm B}T}\frac{\partial^2 f_{\rm m}}{\partial \phi_{\rm m}^2}\bigg|_{\phi_{\rm m}^{I,II}}$ and $D_{\rm m}^{(0)}=\Lambda_{\rm m}^{(0)}k_{\rm B}T$. 
Linearising the binding flux $s$, we find
\begin{equation}
    s=\begin{cases}
    k^{I}(\phi_{\rm m}-\phi_{\rm m}^{I}),\quad x<X_{\rm s}\\
    k^{II}(\phi_{\rm m}-\phi_{\rm m}^{II}),\quad x>X_{\rm s},
    \end{cases}
    \label{eq:si:binding_flux_cases}
\end{equation}
with $k^{I,II} = k_0\left(1-\phi_{\rm m}^{I,II}\right)\left(1-\phi^{I,II}\right)\exp\left[\frac{\mu}{k_{\rm B}T}+\chi_{\rm act}\phi^{I,II}\right]\frac{\nu_{\rm m}}{k_{\rm B}T}\frac{\partial^2 f_{\rm m}}{\partial \phi_{\rm m}^2}\big|_{\phi_{\rm m}^{I,II}}$, where we used a sharp interface model for the bulk with a homogeneous volume fraction $\phi^{I}$ inside and $\phi^{II}$ outside of the droplet. To solve Eq.~\ref{eq:si:stationary} the membrane area fraction $\phi_{\rm m}$ has to read
\begin{equation}
    \phi_{\rm m}=\begin{cases}
    \phi_{\rm m}^{I} + C^{I}\exp\left[-\frac{x-X_{\rm s}}{\lambda^{I}} \right],\quad x<X_{\rm s}\\
    \phi_{\rm m}^{II} + C^{II}\exp\left[-\frac{X_{\rm s}-x}{\lambda^{II}} \right],\quad x>X_{\rm s},
    \end{cases}
    \label{eq:si:area_fraction_cases}
\end{equation}
with the reaction diffusion length scales $\lambda^{I,II}=\sqrt{D_{\rm m}^{I,II}/k^{I,II}}$ and the constants $C^{I,II}$.
When written out, reaction diffusion lengths read
\begin{equation}
    \lambda^{I,II} = \lambda_0\sqrt{ \frac{\phi_{\rm m}^{I,II}}{\left(1-\phi^{I,II}\right)\exp\left[\frac{\mu}{k_{\rm B}T}+\chi_{\rm act}\phi^{I,II}\right]}.}
\end{equation}
Both the membrane area fraction and the membrane flux have to be continuous at $X_{\rm s}$, which implies
\begin{equation}
    \phi_{\rm m}^{I} + C^{I} = \phi_{\rm m}^{II} + C^{II}
    \label{eq:si:area_fraction_continuous}
\end{equation}
and
\begin{equation}
    \frac{D_{\rm m}^{I}}{\lambda^{I}}C^{I} = -\frac{D_{\rm m}^{II}}{\lambda^{II}}C^{II}.
    \label{eq:si:membrane_flux_continuous}
\end{equation}
From Eqs.~\ref{eq:si:area_fraction_continuous} and ~\ref{eq:si:membrane_flux_continuous} we find
\begin{subequations}
\begin{align}
C^{I} &=  -\left(\phi_{\rm m}^{I}-\phi_{\rm m}^{II}\right) \frac{D_{\rm m}^{II}\lambda^{I}}{D_{\rm m}^{II}\lambda^{I}+D_{\rm m}^{I}\lambda^{II}} \\
C^{II} &=  \left(\phi_{\rm m}^{I}-\phi_{\rm m}^{II}\right) \frac{D_{\rm m}^{I}\lambda^{II}}{D_{\rm m}^{II}\lambda^{I}+D_{\rm m}^{I}\lambda^{II}}.
\end{align}
\label{eq:si:C_1_and_C_2}
\end{subequations}
Using Eqs.~\ref{eq:si:binding_flux_cases} and \ref{eq:si:area_fraction_cases} the dipole moment is obtained as
\begin{align}
    \bar{p}&=\int_0^\infty dx\, xs(x)\nonumber\\
    &=k^{I}C^{I}\lambda^{I}\left(X_{\rm s}-\lambda^{I}\right) + k^{II}C^{II}\lambda^{II}\left(X_{\rm s}+\lambda^{II}\right)\nonumber\\
    &=-D_{\rm m}^{I}C^{I}+D_{\rm m}^{II}C^{II},
    \label{eq:si:dipole_full_expression}
\end{align}
where we used Eq.~\ref{eq:si:membrane_flux_continuous} and the relation $k^{I,II}\lambda^{I,II}=D_{\rm m}^{I,II}/\lambda^{I,II}$. To simplify Eq.~\ref{eq:si:dipole_full_expression} further, we note that the area fraction at $X_{\rm s}$ is close to $(\phi_{\rm m}^{I}+\phi_{\rm m}^{II})/2$. The constants $C^{I,II}$ are thus approximated as
\begin{subequations}
\begin{align}
C^{I} &=  -\frac{\phi_{\rm m}^{I}-\phi_{\rm m}^{II}}{2} \\
C^{II} &=  \frac{\phi_{\rm m}^{I}-\phi_{\rm m}^{II}}{2},
\end{align}
\end{subequations}
which simplifies the dipole moment to 
\begin{equation}
    \bar{p} = \left(D_{\rm m}^{I}+D_{\rm m}^{II}\right)\frac{\phi_{\rm m}^{I}-\phi_{\rm m}^{II}}{2}.
\end{equation}
Next, we determine the position of the dipole moment $X_{\rm p}$, with 
\begin{equation}
    X_{\rm p} = \frac{\int_0^\infty dx\,x^2s(x)}{2\int_0^\infty dx\,xs(x)}.
    \label{eq:si:position_dipole_definition}
\end{equation}
Using Eqs.~\ref{eq:si:binding_flux_cases} and \ref{eq:si:area_fraction_cases}, we find
\begin{equation}
    \int_0^\infty dx\,x^2s(x) = 
    -2\left(D_{\rm m}^{I}C^{I}\lambda^{I}+D_{\rm m}^{II}C^{II}\lambda^{II}\right)
    +X_{\rm s}\left(-D_{\rm m}^{I}C^{I}+D_{\rm m}^{II}C^{II}\right)
    -X_{\rm s}^2\left(\frac{D_{\rm m}^{I}}{\lambda^{I}}C^{I}+\frac{D_{\rm m}^{II}}{\lambda^{II}}C^{II}\right),
\end{equation}
where we again use $k^{I,II}\lambda^{I,II}=D_{\rm m}^{I,II}/\lambda^{I,II}$. The position of the dipole moment thus reads
\begin{equation}
    X_{\rm p} = X_{\rm s} + \frac{D_{\rm m}^{I}C^{I}\lambda^{I}+D_{\rm m}^{II}C^{II}\lambda^{II}}{D_{\rm m}^{I}C^{I}-D_{\rm m}^{II}C^{II}}.
    \label{eq:si:position_dipole_step1}
\end{equation}
Using Eq. \ref{eq:si:membrane_flux_continuous}, Eq. \ref{eq:si:position_dipole_step1} simplifies to
\begin{equation}
    X_{\rm p} = X_{\rm s} + \lambda^{I}-\lambda^{II}.
    \label{eq:si:position_dipole_step2}
\end{equation}
The reaction diffusion length scales $\lambda^{I,II}$ depend on both $\phi_{\rm m}^{I,II}$ and $\phi^{I,II}$. Since $\phi_{\rm m}^{I,II}$ varies between 0 and 1 going from largely negative to largely positive $\chi_{\rm act}$ the resulting $\lambda^{I,II}$ change significantly with $\chi_{\rm act}$ as well (Fig.~\ref{fig:length_scales}). 

Together with Eq. \ref{eq:si:shift_Xs-X0} the final expression for the position of the dipole moment becomes
\begin{equation}
    X_{\rm p} = X_0 + \Delta X + \lambda^{I}-\lambda^{II}.
    \label{eq:si:position_dipole_final}
\end{equation}

\subsection{Three-dimensional system}
For the stationary solution, the binding flux $s$ and the radial component of the  membrane flux $j_{\rm m}^{(r)}$ are related as 
\begin{equation}
    s=-\frac{1}{r}\frac{d}{dr}\left(r\, j_{\rm m}^{(r)}\right),
    \label{eq:si:stationary_3d}
\end{equation} 
with $r$ the radial coordinate. Linearizing $j_{\rm m}^{(r)}$ and $s$ around $\phi_{\rm m}^{I}$ and $\phi_{\rm m}^{II}$ respectively, leads to
\begin{equation}
    j_{\rm m}=\begin{cases}
    -D_{\rm m}^{I}\frac{d}{dr}\phi_{\rm m},\quad r<R_{\rm s}\\
    -D_{\rm m}^{II}\frac{d}{dr}\phi_{\rm m},\quad r>R_{\rm s},
    \end{cases}
    \label{eq:si:dipole_membrane_flux_cases_3d}
\end{equation}
and
\begin{equation}
    s=\begin{cases}
    k^{I}(\phi_{\rm m}-\phi_{\rm m}^{I}),\quad r<R_{\rm s}\\
    k^{II}(\phi_{\rm m}-\phi_{\rm m}^{II}),\quad r>R_{\rm s},
    \end{cases}
    \label{eq:si:binding_flux_cases_3d},
\end{equation}
with $R_{\rm s}$ the radial position along the membrane, where the binding flux $s$ is zero and $D_{\rm m}^{I,II}$, $k^{I,II}$ the same diffusion constants and rates as in the two-dimensional case, Eqs.~\ref{eq:si:dipole_membrane_flux_cases}, \ref{eq:si:binding_flux_cases}. Eq.~\ref{eq:si:stationary_3d} is solved by
\begin{equation}
\phi_{\rm m}=\begin{cases}
\phi_{\rm m}^{I}+\overline{C}^{I}\frac{I_0\left( r/\lambda^{I} \right)}{I_0\left( R_{\rm s}/\lambda^{I} \right)};\quad r<R_{\rm s},\\
    \phi_{\rm m}^{II}+\overline{C}^{II}\frac{K_0\left( r/\lambda^{II} \right)}{K_0\left( R_{\rm s}/\lambda^{II} \right)}\quad r>R_{\rm s},
 \end{cases}
\label{eq:si:area_fraction_cases_3d}
\end{equation}
with $I_0$, $K_0$ the modified Bessel function of first and second kind. The constants $\overline{C}^{I,II}$ are determined through the continuity of both the membrane area fraction and the membrane flux at $R_{\rm s}$, which implies
\begin{equation}
    \phi_{\rm m}^{I}+\overline{C}^{I} = \phi_{\rm m}^{II}+\overline{C}^{II} 
    \label{eq:si:area_fraction_continuous_3d}
\end{equation}
and
\begin{equation}
    \frac{D_{\rm m}^{I}}{\lambda^{I}}\overline{C}^{I}\frac{I_1\left( R_{\rm s}/\lambda^{I} \right)}{I_0\left( R_{\rm s}/\lambda^{I} \right)} = -\frac{D_{\rm m}^{II}}{\lambda^{II}}\overline{C}^{II}\frac{K_1\left( R_{\rm s}/\lambda^{II} \right)}{K_0\left( R_{\rm s}/\lambda^{II} \right)}.
    \label{eq:si:membrane_flux_continuous_3d}
\end{equation}
From Eqs.~\ref{eq:si:area_fraction_continuous_3d} and ~\ref{eq:si:membrane_flux_continuous_3d} we find
\begin{subequations}
\begin{align}
\overline{C}^{I} &=  -\left(\phi_{\rm m}^{I}-\phi_{\rm m}^{II}\right)
\frac{D_{\rm m}^{II}\lambda^{I}I_0\left(R_{\rm s}/\lambda^{I}\right)/I_1\left(R_{\rm s}/\lambda^{I}\right)}{D_{\rm m}^{II}\lambda^{I}I_0\left(R_{\rm s}/\lambda^{I}\right)/I_1\left(R_{\rm s}/\lambda^{I}\right) + D_{\rm m}^{I}\lambda^{II}K_0\left(R_{\rm s}/\lambda^{II}\right)/K_1\left(R_{\rm s}/\lambda^{II}\right)} \\
\overline{C}^{II} &=  \left(\phi_{\rm m}^{I}-\phi_{\rm m}^{II}\right)
\frac{D_{\rm m}^{I}\lambda^{II}K_0\left(R_{\rm s}/\lambda^{II}\right)/K_1\left(R_{\rm s}/\lambda^{II}\right)}{D_{\rm m}^{II}\lambda^{I}I_0\left(R_{\rm s}/\lambda^{I}\right)/I_1\left(R_{\rm s}/\lambda^{I}\right) + D_{\rm m}^{I}\lambda^{II}K_0\left(R_{\rm s}/\lambda^{II}\right)/K_1\left(R_{\rm s}/\lambda^{II}\right)}.
\end{align}
\label{eq:si:C_1_and_C_2_3d}
\end{subequations}
We note that the following limits apply 
\begin{subequations}
\label{eq:si:limits} 
\begin{align}
&\lim_{y\to\infty}\frac{I_0(y)}{I_1(y)}=1 \\
&\lim_{y\to\infty}\frac{K_0(y)}{K_1(y)}=1.
\end{align}
\end{subequations}
For large droplets with $R_{\rm s}/\lambda^{I}\gg1$, $R_{\rm s}/\lambda^{II}\gg1$ the constants $\overline{C}^{I,II}$ in a three-dimensional system are equal to the constants $C^{I,II}$ in a two-dimensional system.\\
To determine the line dipole moment, it is useful to recapitulate two further mathematical relations involving the modified Bessel function of first and second kind:
\begin{subequations}
\begin{equation}
    \int_0^ydx\, x^2I_0(x)=y^2I_1(y)+\frac{\pi}{2}yI_1(y)\left[\frac{I_0(y)}{I_1(y)}L_1(y) -L_0(y)\right],
\end{equation}
\begin{equation}
    \int_y^\infty dx\, x^2K_0(x)=\frac{\pi}{2}+y^2K_1(y)-\frac{\pi}{2}yK_1(y)\left[\frac{K_0(y)}{K_1(y)}L_1(y) +L_0(y)\right],
\end{equation}
\end{subequations}
with $L_{\rm \alpha}(y)$ the modified Struve function. Using the limits
\begin{subequations}
\begin{equation}
    \lim_{y\to\infty}\frac{I_0(y)}{I_1(y)}L_1(y) -L_0(y)=-\frac{2}{\pi},
\end{equation}
\begin{equation}
    \lim_{y\to\infty}\frac{1}{yK_1(y)}-\frac{K_0(y)}{K_1(y)}L_1(y) -L_0(y)=\frac{2}{\pi},
\end{equation}
\end{subequations}
we obtain the asymptotic behavior
\begin{subequations}
\label{eq:si:asymptotic}
\begin{equation}
    \int_0^ydx\, x^2I_0(x)\approx(y^2-y)I_1(y),\text{ for }y\gg1,
\end{equation}
\begin{equation}
    \int_y^\infty dx\, x^2K_0(x)\approx(y^2+y)K_1(y),\text{ for }y\gg1.
\end{equation}
\end{subequations}
Using Eqs.~\ref{eq:si:binding_flux_cases_3d}, \ref{eq:si:area_fraction_cases_3d} and \ref{eq:si:asymptotic} the line dipole moment for large droplets ($R_{\rm s}/\lambda^{I}\gg1$, $R_{\rm s}/\lambda^{II}\gg1$) is obtained as
\begin{align}
    \bar{p}_{L}&=\int_0^\infty dr\, r^2s(r)\nonumber\\
    &= k^{I}\overline{C}^{I}\frac{I_1\left(R_{\rm s}/\lambda^{I}\right)}{I_0\left(R_{\rm s}/\lambda^{I}\right)}\left(R_{\rm s}^2\lambda^{I}-R_{\rm s}\left(\lambda^{I}\right)^2\right)+
    k^{II}\overline{C}^{II}\frac{K_1\left(R_{\rm s}/\lambda^{II}\right)}{K_0\left(R_{\rm s}/\lambda^{II}\right)}\left(R_{\rm s}^2\lambda^{II}-R_{\rm s}\left(\lambda^{II}\right)^2\right)\nonumber\\
    &= \left(D_{\rm m}^{II}\overline{C}^{II}\frac{K_1\left(R_{\rm s}/\lambda^{II}\right)}{K_0\left(R_{\rm s}/\lambda^{II}\right)}-D_{\rm m}^{I}\overline{C}^{I}\frac{I_1\left(R_{\rm s}/\lambda^{I}\right)}{I_0\left(R_{\rm s}/\lambda^{I}\right)}\right)R_{\rm s}
    \label{eq:si:dipole_full_expression_3d}
\end{align}
where we used Eq.~\ref{eq:si:membrane_flux_continuous_3d} and the relation $k^{I,II}\lambda^{I,II}=D_{\rm m}^{I,II}/\lambda^{I,II}$.\\
Using Eq.~\ref{eq:si:limits}, Eq.~\ref{eq:si:dipole_full_expression_3d} is further simplified 
\begin{equation}
    \bar{p}_{L}=\left(D_{\rm m}^{II}C^{II}-D_{\rm m}^{I}C^{I}\right)R_{\rm s},
\end{equation}
with $C^{I,II}$ given in Eq.~\ref{eq:si:C_1_and_C_2}. 
Comparing the dipole moment $\bar{p}$ in a two-dimensional system, Eq.~\ref{eq:si:dipole_full_expression} and the line dipole moment $\bar{p}_{\rm L}$ in a three-dimensional system, we find
\begin{equation}
    \bar{p}_{\rm L} = R_{\rm s}\bar{p}.
\end{equation}
Next, we determine the position of the dipole moment $R_{\rm p}$, with 
\begin{equation}
    R_{\rm p} = \frac{\int_0^\infty dr\,r^3s(r)}{2\int_0^\infty dr\,r^2s(r)}.
    \label{eq:si:position_dipole_definition_3d}
\end{equation}
We use the relations
\begin{subequations}
\begin{equation}
    \int_0^ydx\, x^3I_0(x)=\left(y^3+4y\right)I_1(y)-2y^2I_0(y),
\end{equation}
\begin{equation}
    \int_y^\infty dx\, x^3K_0(x)=\left(y^3+4y\right)K_1(y)+2y^2K_0(y),
\end{equation}
\end{subequations}
to evaluate the integral
\begin{align}
    \int_0^\infty dr\, r^3s(r)&=
    k^{I}\overline{C}^{I}\left[\left(R_{\rm s}^3\lambda^{I}+4R_{\rm s}(\lambda^{I})^3\right)\frac{I_1(R_{\rm s}/\lambda^{I})}{I_0(R_{\rm s}/\lambda^{I})} - 2R_{\rm s}^2(\lambda^{I})^2\right] +
    k^{II}\overline{C}^{II}\left[\left(R_{\rm s}^3\lambda^{II}+4R_{\rm s}(\lambda^{II})^3\right)\frac{K_1(R_{\rm s}/\lambda^{II})}{K_0(R_{\rm s}/\lambda^{II})} + 2R_{\rm s}^2(\lambda^{II})^2\right]
    \nonumber\\
    &= 4R_{\rm s}\left(D_{\rm m}^{I}\lambda^{I}\overline{C}^{I}\frac{I_1(R_{\rm s}/\lambda^{I})}{I_0(R_{\rm s}/\lambda^{I})} + D_{\rm m}^{II}\lambda^{II}\overline{C}^{II}\frac{K_1(R_{\rm s}/\lambda^{II})}{K_0(R_{\rm s}/\lambda^{II})}\right)+2R^2_{\rm s}\left(D_{\rm m}^{II}\overline{C}^{II}-D_{\rm m}^{I}\overline{C}^{I}\right),
    \label{eq:si:dipole_position_step1_3d}
\end{align}
where we used Eq.~\ref{eq:si:membrane_flux_continuous_3d} and the relation $k^{I,II}\lambda^{I,II}=D_{\rm m}^{I,II}/\lambda^{I,II}$. For large droplets, with $R_{\rm s}/\lambda^{I},R_{\rm s}/\lambda^{II}\gg1$ Eq.~\ref{eq:si:dipole_position_step1_3d} simplifies to
\begin{equation}
    \int_0^\infty dr\, r^3s(r) = 4R_{\rm s}\left(D_{\rm m}^{I}\lambda^{I}C^{I}+D_{\rm m}^{II}\lambda^{II}C^{II}\right)+2\left(D_{\rm m}^{II}C^{II}-D_{\rm m}^{I}C^{I}\right)R_{\rm s}^2.
\end{equation}
Thus the position of the dipole moment reads
\begin{equation}
    R_{\rm p}=R_{\rm s}+2\left(\lambda^{I}-\lambda^{II}\right).
\end{equation}
Together with Eq. \ref{eq:si:shift_Xs-X0} the final expression for the position of the dipole moment becomes
\begin{equation}
    R_{\rm p} = R_0 + \frac{ \frac{\phi^{I}+\phi^{II}}{2} - \nicefrac{\left(\frac{\nu_{\rm m}}{k_{\rm B}T}\frac{\partial f_{\rm m}}{\partial \phi_{\rm m}}\big|_{\frac{\phi_{\rm m}^{I}+\phi_{\rm m}^{II}}{2}}-\frac{\mu}{k_{\rm B}T}\right)}{\chi_{\rm act}}}{\sqrt{
    \frac{2}{\kappa}
    \left( f(\phi_{\frac{1}{2}}) - f(\phi^{II}) -\frac{\mu}{\nu}(\phi_{\frac{1}{2}}-\phi^{II})
    \right)}} + 2\left(\lambda^{I}-\lambda^{II}\right),
    \label{eq:si:position_dipole_final_3d}
\end{equation}
with $R_0$ the base radius of the droplet.

\bibliography{supplement}